\documentclass{ametsoc}

\journal{jpo}

\bibpunct{(}{)}{;}{a}{}{,}

\title{On stratification, barotropic tides, and secular changes in surface tidal elevations: Two-layer analytical model}

\nolinenumbers

\authors{
Alfredo N. Wetzel
\correspondingauthor{ Alfredo N. Wetzel, Department of Mathematics, 5080 East Hall, 530 Church Street, University of Michigan, Ann Arbor, MI, 48109-1043.}
}

\affiliation{Applied and Interdisciplinary Mathematics Program, Department of Mathematics, University of Michigan, Ann Arbor, Michigan}

\email{wreagan@umich.edu}

\extraauthor{Brian K. Arbic}
\extraaffil{Department of Earth and Environmental Sciences, University of Michigan, Ann Arbor, Michigan}

\extraauthor{Ivana Cerovecki and Myrl C. Hendershott}
\extraaffil{Physical Oceanography Research Division, Scripps Institution of Oceanography, University of California, San Diego, California}

\extraauthor{Richard H. Karsten}
\extraaffil{Department of Mathematics and Statistics, Acadia University, Wolfville, Nova Scotia, Canada}

\extraauthor{Peter D. Miller}
\extraaffil{Department of Mathematics, University of Michigan, Ann Arbor, Michigan}

\extraauthor{Joseph F. Molinari}
\extraaffil{Department of Mathematics, Florida State University, Tallahassee, Florida}

\abstract{
In this study the influence of stratification on surface tidal elevations in a two-layer analytical model is examined. The model assumes linearized, non-rotating, shallow-water dynamics in one dimension with astronomical forcing and allows for arbitrary topography. Using a natural modal separation, both large scale (barotropic) and small scale (baroclinic) components of the surface tidal elevation are shown to be comparably affected by stratification. It is also shown that the topography and basin boundaries affect the sensitivity of the barotropic surface tide to stratification significantly. This paper, therefore, provides a framework to understand how the presence of stratification impacts barotropic as well as baroclinic tides, and how climatic perturbations to oceanic stratification contribute to secular variations in tides.
Results from a realistic-domain global numerical two-layer tide model are briefly examined and found to be qualitatively consistent with the analytical model results.
}

\begin{document}

%% Necessary!
\maketitle

%%%%%%%%%%%%%%%%%%%%%%%%%%%%%%%%%%%%%%%%%%%%%%%%%%%%%%%%%%%%%%%%%%%%%
% MAIN BODY OF PAPER
%%%%%%%%%%%%%%%%%%%%%%%%%%%%%%%%%%%%%%%%%%%%%%%%%%%%%%%%%%%%%%%%%%%%%

\section{Introduction}

In this paper we utilize an analytical model to examine the impact of deep ocean stratification and climatic perturbations to this stratification on both large-horizontal scale (barotropic) and small-horizontal scale (baroclinic) components of the surface tidal elevations. 
Numerous studies have determined that, at specific locations, surface tidal elevations  have undergone secular changes over the last century
\citep[e.g.,][]{Cartwright1971, Woodworth+al1991, Flick+al2003, Colosi+Munk2006, Ray2006, Ray2009, Jay2009, Woodworth2010, Muller+al2011, Zaron+Jay2014}. 
One motivation for this work arises from the possibility that these observed secular signals are caused in part by changes in the oceanic stratification.

Another motivation for this work is to explain why the addition of stratification
alters the large-horizontal scale (barotropic) tide in realistic-domain global numerical simulations
at the same time that, as is well known, it introduces a
 small-scale (baroclinic) tide.
\cite{Arbic+al2004} briefly noted
that the dominant large amplitude, large-horizontal scale (barotropic) tide in a two-layer global model is
distinct from the barotropic tide in a one-layer global model; see Figure 10 in that paper. 
The large-scale amplitudes of the one- and two-layer results differ by roughly 10$\%$, while the phases differ by about 10$^{\circ}$. The shift in barotropic tides with the addition of stratification into a realistic-domain global model can also be seen in Figure 1 of \cite{Shriver+al2014}.
In this paper we investigate whether a simple analytical model can produce qualitatively similar results as seen in these global numerical simulations run in realistic domains. 

Our study is global in nature.  Related regional studies include \cite{Kang+al2002}, who examined seasonal changes in stratification and tides in the Yellow and East China Seas, and \cite{Colosi+Munk2006}, who examined the impact of internal tides on the secular variations in surface tidal elevations recorded in tide gauges.  Another study related to the present one is \cite{Muller2012}, who examined the effects of seasonally varying coastal stratifications on the tides using an idealized model.

The underlying mechanism examined here for the impact of stratification on barotropic tides is that stratification introduces a perturbation to barotropic gravity wave speeds \citep{Gill1982}.  This mechanism for the impact of stratification on barotropic tides differs from the mechanism examined by \cite{Muller2012}, which involves eddy viscosities rather than gravity wave speeds and is oriented towards the shelves rather than the deep ocean.

The analytical model used in this paper assumes two-layer, linearized, non-rotating, shallow-water dynamics in one horizontal dimension. We assume astronomical forcing and allow for arbitrary  bottom topography in both finite and infinite basin geometries. For convenience, all numerical results in this paper utilize a topography consisting of a Gaussian bump in the center of the basin. 
Baroclinic tides are generated by barotropic tidal flow over the topographic feature.
We impose no-normal flow boundary conditions at the basin boundaries for a finite basin, or equivalently decay to zero at infinity for the infinite basin.
We obtain model solutions using both a Fourier series and a Neumann series expansion.
Importantly, the Neumann series contains built in barotropic and baroclinic modes allowing both the velocities and elevations to be naturally decomposed.

We examine the effects of stratification on the surface elevations primarily by computing the surface elevations for different stratification parameters on both finite and infinite basin geometries. Quite visible in these results, at least for the finite basin, is a measurable change in the barotropic as well as baroclinic components of the surface elevation, brought about by changes in stratification. In addition, we find that the effects of stratification on the barotropic component of surface elevation are significantly influenced by the topography and basin boundaries.

%%%%%%%%%%%%%%%%%%%%%

\section{Illustrative global numerical simulations}\label{sec:numerical}

To illustrate the effects of stratification on barotropic tides, and to motivate our analytical model, we briefly discuss numerical one- and two-layer tide simulations performed in a realistic near-global domain. The simulations are executed with the Hallberg Isopycnal Model  \citep[HIM;][]{Hallberg+Rhines1996} on a $1/8^{\circ}$ grid  and a multiplicative topographic wave drag factor of $5$, where the wave drag acts only in waters deeper than $1000$ m. 
This topographic drag factor is chosen to minimize the discrepancy in deep-ocean tidal elevations in the tide model and the GOT99 satellite-altimeter constrained tide model \citep{Ray1999} and yields a 6.69 cm error in our one-layer 1/8$^{\circ}$ run. 
 As in \cite{Arbic+al2004}, the wave drag acts only on the bottom flow in our two-layer simulations.  
For simplicity, we apply only astronomical forcing for $M_2$, the principal lunar semidiurnal tide. 
The self-attraction and loading term is computed iteratively using the full spherical harmonic treatment \citep{Hendershott1972}.
We refer the interested reader to \cite{Arbic+al2004} for further details on the model setup and parameters.

We evaluate the strength of the baroclinic tides in the 1/8$^{\circ}$ HIM results utilized in this paper via the same methods used to evaluate the 1/12$^{\circ}$ HYCOM (HYbrid Coordinate Ocean Model) results in \cite{Arbic+al2010}.  In the HIM ``interface perturbation'' run to be described shortly, we find that the root mean square (RMS) $M_{2}$ internal tide perturbations to sea surface elevations, averaged over a large box around Hawai'i 35 degrees in latitude by 50 degrees in longitude, are 0.64 cm for amplitude and 3.56 degrees for phase.  These numbers compare well to the 0.87 cm and 4.35 degree perturbations estimated in the same box from the along-track satellite altimeter observations of \cite{Ray+Mitchum1996, Ray+Mitchum1997}.  The 1/8$^{\circ}$ ``$g'$ perturbation'' and ``control'' runs to be described shortly yield very similar numbers.  Therefore the 1/8$^{\circ}$ two-layer HIM simulations produce internal tides that are of comparable magnitudes to those in observations.

\begin{figure}[h!]
\begin{center}
\includegraphics[scale=0.68]{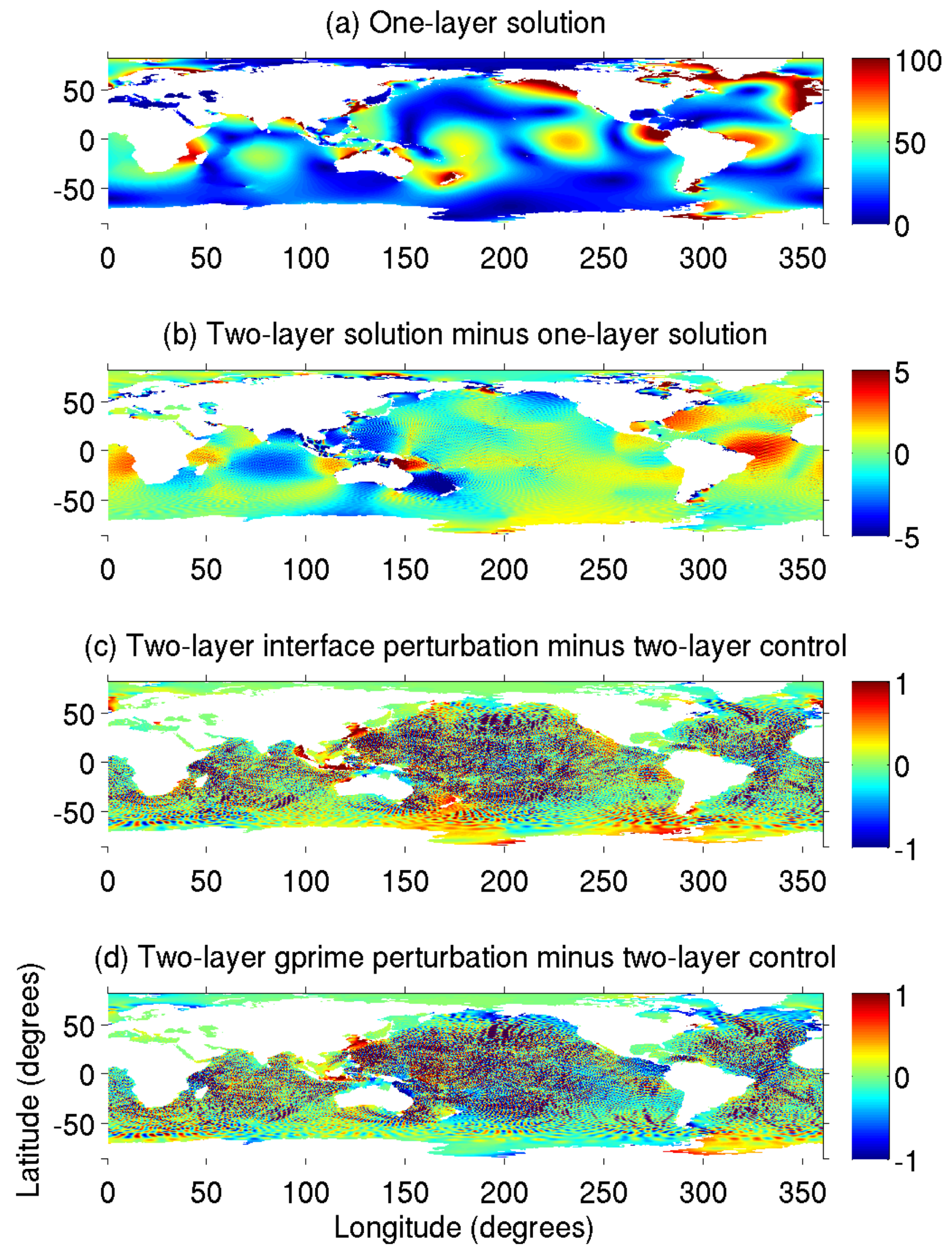}
\caption{Surface elevation amplitudes (cm) in 1/8$^{\circ}$ $M_{2}$ simulations:  (a) one-layer control simulation, (b) two-layer control simulation minus one-layer control simulation, (c) two-layer ``interface perturbation'' simulation (with layer interface at 800 m) minus two-layer control simulation (with layer interface at 700 m), (d) two-layer ``$g'$ perturbation'' simulation with perturbed $g'$ value of $1.78 \times 10^{-2}$ m s$^{-2}$ minus two-layer simulation with control $g'$ value of $1.64 \times 10^{-2}$ m s$^{-2}$.} \label{fig:global_AMP} % The fourth iteration of the self-attraction and loading (SAL) term is used for each plot.}
\end{center}
\end{figure}

\begin{figure}[h!]
\begin{center}
\includegraphics[scale=0.68]{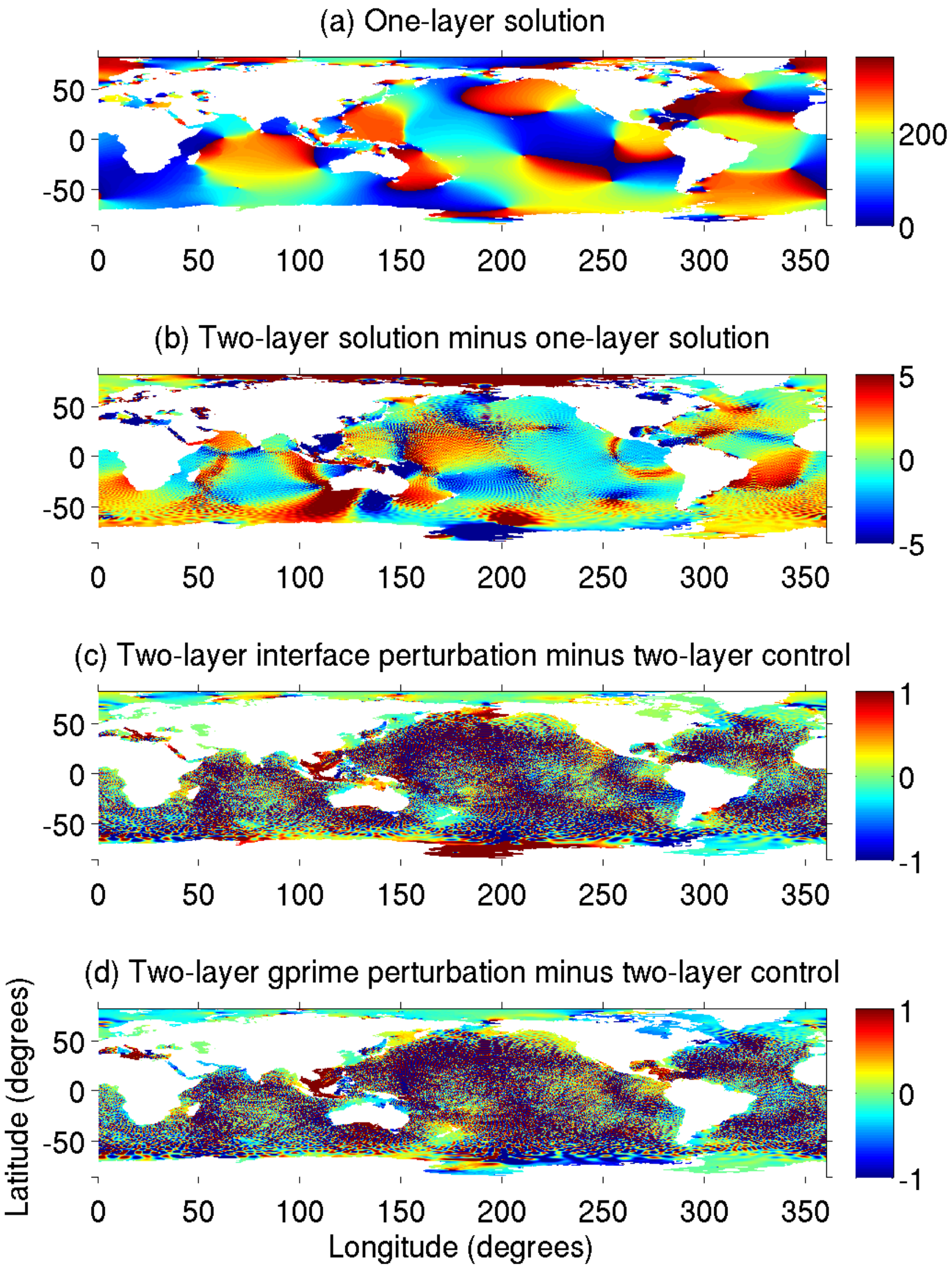}
\caption{As in Figure \ref{fig:global_AMP} but for phase of surface elevations (degrees) in 1/8$^{\circ}$ $M_{2}$ simulations.}\label{fig:global_PHASE}
\end{center}
\end{figure}

Figures \ref{fig:global_AMP} and \ref{fig:global_PHASE} display the impacts of stratification, and of perturbations to stratification, on the tidal amplitude and phase. Figure \ref{fig:global_AMP}(a) displays the global M$_{2}$ surface elevation amplitude in our 1/8$^{\circ}$ one-layer control simulation. 
Figure \ref{fig:global_AMP}(b) displays the difference in surface elevation amplitudes between the 1/8$^{\circ}$ two-layer control simulation, having an interface at $700$ m and reduced gravity of $g'=1.64 \times 10^{-2}$ m s$^{-2}$, and the one-layer control simulation.
Differences between the two- and one-layer simulations are clearly seen in the large-scale barotropic tides as well as in the introduction of small-scale barotropic tides.
A perturbation to the stratification in the two-layer numerical model, either in the form of an interface lying at 800 m rather than 700 m, or in an increase in $g'$ to $1.78 \times 10^{-2}$ m s$^{-2}$, leads to further alterations in both the large-scale barotropic and small-scale baroclinic tides; see Figures \ref{fig:global_AMP}(c) and (d).  Our choices of reduced gravity $g'$ and interface values are explained in section \ref{sec:num} \ref{sec:params}.
Figure \ref{fig:global_PHASE} demonstrates that changes also take place in phase, when first stratification and second climatically perturbed stratifications are introduced into the model. 
Figures \ref{fig:global_AMP} and \ref{fig:global_PHASE} demonstrate that the change in the barotropic tide due to stratification is a robust effect. A more detailed examination of the numerical results, and comparison to observed secular changes, is reserved for a paper in preparation. 

%%% Governing Equations

\section{Governing equations}

\begin{figure}[h!]
\begin{center}
\includegraphics[scale=0.55]{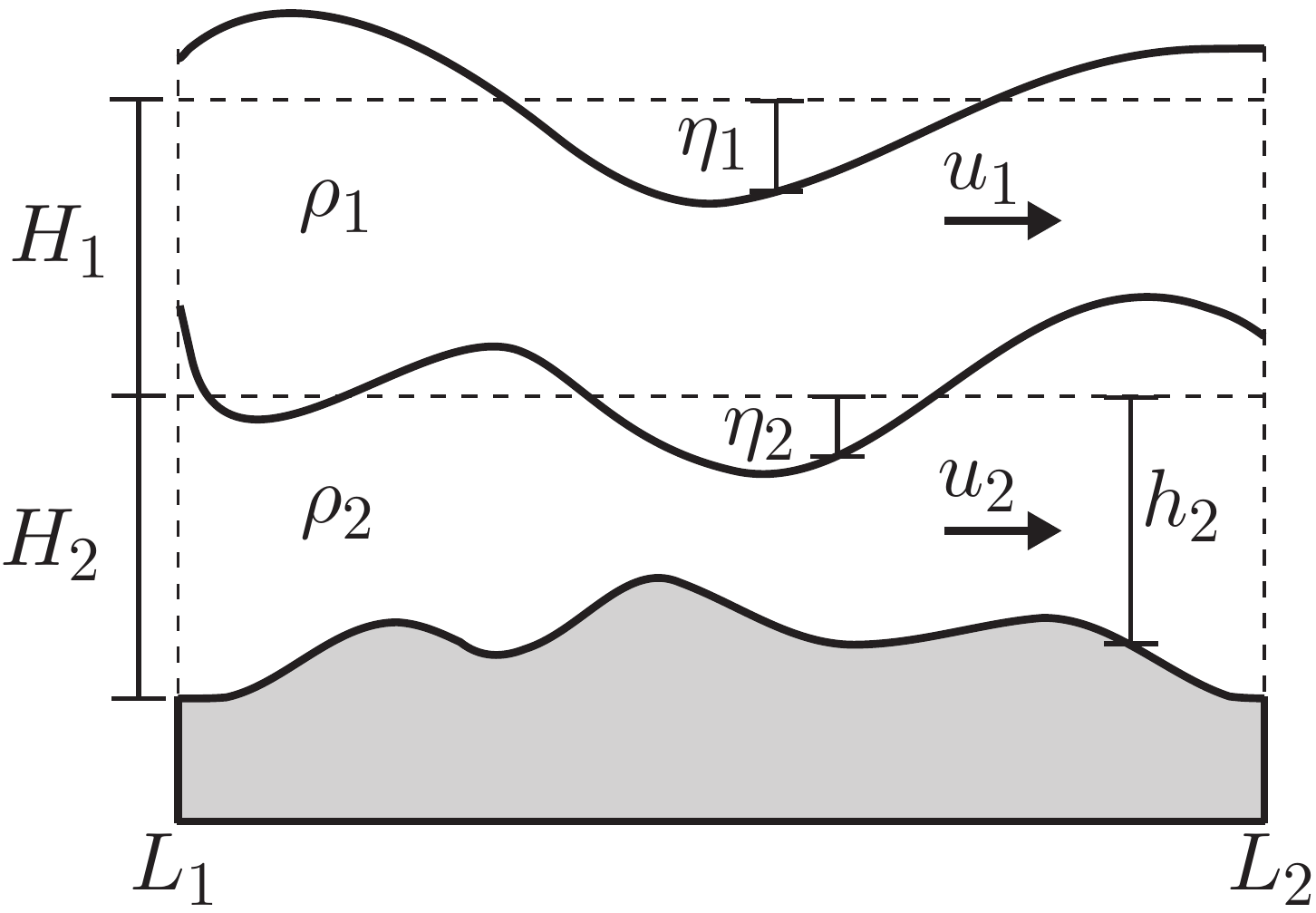}
\caption{Sketch of analytical two-layer model.}\label{fig:sketch}
\end{center}
\end{figure}

Let $u_1(x,t)$, $u_2(x,t)$, $\eta_1(x,t)$, $\eta_2(x,t)$ be, respectively, the upper and lower layer velocities and the perturbation surface and internal elevations (displacements), where $x$ denotes the spatial coordinate and $t$ denotes time. Let the resting layer depths be $H_{1}$ and $h_{2}(x)$, where subscripts 1 and 2 denote the upper and lower layers, respectively. Assuming linearized, non-rotating, shallow-water dynamics in one dimension with astronomical forcing yields the upper and lower layer mass conservation equations
\begin{equation}
\frac{\partial}{\partial t} (\eta_{1}-\eta_{2}) + H_{1} \frac{\partial u_{1}}{\partial x}=0,
\end{equation}
\begin{equation}
\frac{\partial \eta_{2}}{\partial t} + \frac{\partial }{\partial x} \left[ h_{2}(x) u_{2} \right]=0,
\end{equation}
respectively, and the upper and lower layer momentum equations
\begin{equation}
\frac{\partial u_{1}}{\partial t} = -g \frac{\partial \eta_{1}}{\partial x} + g \frac{\partial}{\partial x} \eta_0 e^{i(kx + \omega t)}- r_{1}u_{1},
\end{equation}
\begin{equation}
\frac{\partial u_{2}}{\partial t} = \left(g'-g\right) \frac{\partial \eta_{1}}{\partial x} -g' \frac{\partial \eta_{2}}{\partial x} +g \frac{\partial}{\partial x} \eta_0 e^{i(kx + \omega t)}  -r_{2} u_{2}, \label{eqref:etap}
\end{equation}
respectively. Here $g$ denotes gravity, $g'=g(\rho_{2}-\rho_{1})/\rho_{2}$ is the reduced gravity with $\rho_1$ and $\rho_2$ being the upper and lower layer densities (note $\rho_1 \leq \rho_2$), $r_1$ and $r_{2}$ are damping rates on the upper and lower layer flows, respectively, and the astronomical forcing is given by $\eta_0 e^{i (k x +\omega t)}$, where $\eta_0$, $k$, and $\omega$ are the amplitude, wavenumber, and frequency of the forcing. A sketch of the model is given in Figure \ref{fig:sketch}.

We impose the no-normal flow boundary conditions
\begin{equation}
u_1(L_1,t) = u_1(L_2,t)=u_2(L_1,t)=u_2(L_2,t) =0, \label{eqn:bd_cnd}
\end{equation}
where $L_1$ and $L_2$ are the basin boundaries. For an infinite basin we impose a decay condition on $u_1(x,t)$ and $u_2(x,t)$; a modification of \eqref{eqn:bd_cnd} with the understanding that $L_1 \to -\infty$ and $L_2 \to \infty$.

We non-dimensionalize the governing equations using:
\begin{equation}
x=Lx^{*}, \quad 
t=t^{*}/\omega, \quad
\eta_{j}=\eta_{0} \eta_{j}^{*}, \quad 
u_{j}=\eta_{0} \sqrt{\frac{g}{H_{1}+H_{2}}} u_{j}^{*}, \quad
h_{2}=H_{2} [1 - \sigma b^*(x^*)],
\end{equation}
where the asterisks denote non-dimensional variables, and $j = 1,2$. The parameter $L$ is a characteristic length of the system, e.g. the basin size in a finite basin, or a topographical length scale in an infinite basin. The quantity $\sigma b^*(x^*)$ is a non-dimensionalization of the bottom topography such that $0 \leq b^*< 1$ and $\sigma$ is a scaling parameter. Hence, the quantity $H_2$ is the maximum value attained by $h_2$ in the given domain.

The non-dimensional governing equations, after dropping the
asterisks, are
\begin{equation}
\frac{\partial}{\partial t} (\eta_{1}-\eta_{2}) + \epsilon \gamma \frac{\partial u_{1}}{\partial x}=0,\label{mass1}
\end{equation}
\begin{equation}
\frac{\partial \eta_{2}}{\partial t} + \epsilon (1-\gamma) \frac{\partial}{\partial x} \left[(1 - \sigma b) u_2\right]=0, \label{mass2}
\end{equation}
\begin{equation}
\frac{\partial u_{1}}{\partial t} + \epsilon \frac{\partial \eta_{1}}{\partial x} + \delta_{1} u_{1} = \epsilon \frac{\partial}{\partial x} \left[ e^{i (\phi x + t)} \right], \label{mom1}
\end{equation}
\begin{equation}
\frac{\partial u_{2}}{\partial t} + \epsilon \frac{\partial}{\partial x} \left[ (1 - \alpha) \eta_{1}+\alpha \eta_{2} \right] + \delta_{2} u_{2} = \epsilon \frac{\partial}{\partial x} \left[ e^{i (\phi x + t)} \right], \label{mom2}
\end{equation}
where the governing non-dimensional parameters are
\begin{equation}
\epsilon = \frac{\sqrt{g(H_{1}+H_{2})}}{\omega L}, \quad
\gamma=\frac{H_{1}}{H_{1}+H_{2}}, \quad
\alpha=\frac{g'}{g}, \quad
\delta_{1}=\frac{r_{1}}{\omega}, \quad  \delta_{2}=\frac{r_{2}}{\omega}, \quad \phi = kL.
\end{equation}
Thus, $\delta_1$, $\delta_2$ are the damping parameters and $\alpha$, $\gamma$ are the two stratification parameters. By definition $0 \leq \alpha < 1$, where $\alpha = 0$ represents unstratified flow. Similarly, by definition, $0 \leq \gamma \leq 1$. The parameters $\epsilon$ and $\phi $ are measures of the forcing period and wavenumber. The boundary conditions simply become
\begin{equation}
u_1(\ell_1,t) = u_1(\ell_2,t)=u_2(\ell_1,t)=u(\ell_2,t) =0,
\end{equation}
where $\ell_1 = L_1/L$ and $\ell_2 = L_2/L$. 

In the open ocean, since tidal flows are relatively weak, the quadratic bottom boundary layer drag is also weak \citep{Munk1997}. We assume that bottom drag in the open ocean is dominated by topographic internal wave drag \citep[among others]{Jayne+StLaurent2001, Arbic+al2004, Egbert+al2004}. In our treatment we allow drag on the upper layer flow as a convenience that proves useful in examining limiting cases; we believe that the bottom drag-only case ($\delta_1 =0$) is more relevant for the ocean. We note also that the $g'$ term in front of the derivative of $\eta_1$ in equation \eqref{eqref:etap} arises from the fact that we do not make the Boussinesq approximation when deriving these equations. It is worth noting that had we assumed the Boussinesq approximation, the effects of stratification on the barotropic and baroclinic surface tides would be qualitatively similar. The differences between Boussinesq and non-Boussinesq solutions, however, will not be pursued further in this paper.

Note that, in the absence of topography, setting $\alpha=0$ and $\delta_{1}=\delta_{2}$ yields the governing equations in the one-layer shallow water model used in \cite{Arbic+al2009}. This follows from the fact that if we set $\alpha=0$ (so that the layer densities are equal) and $\delta_{1}=\delta_{2}$ (equal damping rates in the two layers), then the momentum equations for both $u_{1}$ and $u_{2}$ are identical to the momentum equation in the one-layer case.

%%%%%% Finite Domain Model % Analytical solution

\section{Analytical solution}

To solve the two-layer system \eqref{mass1}$-$\eqref{mom2}, we assume a separable solution of the form
\begin{equation}
u_1=U_1(x) e^{it}, \quad u_2=U_2(x) e^{it}, \quad \eta_1=N_1(x) e^{it}, \quad \eta_2=N_2(x) e^{it}. \label{separation}
\end{equation}
This recovers the steady-state solution, which is stable provided there is damping. Substituting these into the mass equations \eqref{mass1}, \eqref{mass2} and solving for $N_1$, $N_2$ gives the system
\begin{equation}
N_1(x)=i\epsilon \gamma U_1' + i\epsilon (1-\gamma) \left[ \left(1 - \sigma b\right) U_2 \right]', \label{mass11}
\end{equation}
\begin{equation}
N_2(x)=i\epsilon (1-\gamma)  \left[ \left( 1 - \sigma b\right) U_2 \right]' . \label{mass22}
\end{equation}
Except for $g'$, prime notation denotes derivatives of a function with respect to $x$ here and in the rest of the paper. This system can be written concisely in matrix form as
\begin{equation}
\textbf{N}(x) =i \epsilon \left( \mathbb{I} + \textbf{e}_{12} \right) \boldsymbol{\Gamma}\textbf{U}'(x), \label{Neq}
\end{equation}
where $\textbf{N}(x) =\left[\begin{smallmatrix} N_1\\ N_2 \end{smallmatrix} \right]$, $\mathbb{I} = \left[\begin{smallmatrix} 1&0\\ 0&1 \end{smallmatrix} \right]$ is the identity matrix, $\textbf{e}_{12} = \left[\begin{smallmatrix} 0 & 1\\ 0&0 \end{smallmatrix} \right]$, $\boldsymbol{\Gamma} = \left[\begin{smallmatrix} \gamma & 0 \\ 0 & 1 -\gamma \end{smallmatrix} \right]$, and
$\textbf{U}(x) =  (\mathbb{I} - \sigma b \textbf{e}_{22}) \left[\begin{smallmatrix} U_1\\ U_2 \end{smallmatrix} \right] =  \left[\begin{smallmatrix} U_1\\ (1 - \sigma b)U_2 \end{smallmatrix} \right]$ with $\textbf{e}_{22} = \left[\begin{smallmatrix} 0 &0\\ 0&1 \end{smallmatrix} \right]$. Similarly, with the use of equations \eqref{mass11} and \eqref{mass22}, substituting \eqref{separation} into the momentum equations \eqref{mom1} and \eqref{mom2} gives
\begin{equation}
s_1 U_{1} + U_{1}'' + \left(\frac{1-\gamma}{\gamma}\right) \left[ \left( 1 - \sigma b \right) U_2 \right]'' 
= \frac{\phi}{\epsilon \gamma}  e^{i\phi x}, \label{mom11}
\end{equation}
\begin{equation}
s_2 U_{2} + (1-\alpha) \left(\frac{\gamma}{1 - \gamma}\right) U_{1}''+ \left[ \left(1 - \sigma b \right) U_2 \right]''
= \frac{\phi}{\epsilon(1 - \gamma)} e^{i\phi x}, \label{mom22}
\end{equation}
where $\displaystyle s_1 = \frac{1 - i \delta_1}{\epsilon^2 \gamma}$ and $\displaystyle s_2 = \frac{1 - i \delta_2}{\epsilon^2 (1-\gamma)}$. Multiplying \eqref{mom11} by $s_2$ and \eqref{mom22} by $s_1$ allows us to write these equations as
\begin{equation}
\textbf{A} \textbf{U}(x) + \sigma \textbf{A} \textbf{e}_{22} \left( \frac{b(x)}{1 - \sigma b(x)} \right) \textbf{U}(x) + \textbf{B} \textbf{U}''(x) = \textbf{F}(x), \label{Umatsystem}
\end{equation}
where
\begin{equation}
\textbf{A} = s_1 s_2 \mathbb{I}, \quad \textbf{B} = \begin{bmatrix} s_2 & s_2 \left( \frac{1 - \gamma}{\gamma} \right) \\ s_1 (1 - \alpha) \left( \frac{\gamma}{1-\gamma} \right) & s_1 \end{bmatrix}\text{, and} \quad \textbf{F}(x) = \frac{\phi}{\epsilon} \begin{bmatrix} \frac{s_2}{\gamma} \\ \frac{s_1}{1-\gamma}\end{bmatrix}   e^{i \phi x}.
\end{equation}

In the subsequent subsections we present three distinct methods, each having different strengths, to solve this set of equations. 
First, we show the derivation of a Fourier series solution.
This solution method is limited to the finite basin problem, but is simple to implement and is robust numerically. 
Second, we introduce a Neumann series solution. 
The Neumann series solution is numerically sensitive to the topography, but suggests an analytically valuable decomposition into barotropic and baroclinic modes.
Lastly, we introduce a scattering solution to the problem. 
We include the scattering solution method here because it bears similarities to the internal tide generation problem of great recent interest to the community \citep[e.g.,][]{Llewellyn+al2001, StLaurent+al2003, Khatiwala2003}.
The scattering solution is, however, not emphasized in this paper as much as the other two methods because, first, the finite basin is more realistic than the infinite basin, and second, the scattering solution still requires prior knowledge of the ``incident'' barotropic tidal velocity on the topography;
 whereas we are interested in obtaining, without prior assumptions, the full barotropic and baroclinic tidal solution. 

%%%%%%%%%%%%%%%%%%%

\subsection{Fourier series solution}

In the finite basin case, we can construct a solution for general topography using Fourier series. Here we non-dimensionalize using the basin width, making $\ell_1 = -1, \ell_2=0$. To satisfy the boundary conditions the velocities must be expansions of sines only, i.e,
\begin{equation}
U_1(x) =\sum_{n=1}^{\infty} c_n \sin n\pi x, \;\; U_2(x) =\sum_{n=1}^{\infty} d_n \sin n\pi x, \;\; (1-\sigma b(x))U_2(x) =\sum_{n=1}^{\infty} e_n \sin n\pi x,
\end{equation}
where
\begin{equation}
e_n = 2 \int_{-1}^0 \left(1 - \sigma b(x)\right) U_2(x) \sin (n\pi x) \, dx = d_n - \sum_{m=1}^{\infty}  d_m \sigma_{mn}
\end{equation}
with
\begin{equation}
\sigma_{mn} = 2 \sigma \int_{-1}^{0}  b(x) \sin (m\pi x) \sin (n\pi x) \, dx.
\end{equation}
We may also write the forcing term in a Fourier series expansion. That is,
\begin{equation}
e^{i\phi x} =\sum_{n=1}^{\infty} f_n \sin n\pi x,
\end{equation}
where
\begin{equation}
f_n = 2 \int_{-1}^{0} e^{i \phi x} \sin (n \pi x) \; dx = \frac{2n\pi[1-(-1)^n e^{-i\phi}]}{\phi^2-(n\pi)^2}.
\end{equation}
Therefore, equations \eqref{mom11} and \eqref{mom22} can be written in terms of the coefficients $c_n$ and $d_n$ as
\begin{equation}
\left( (n \pi)^2 - s_1 \right) c_n + (n \pi)^2  \left(\frac{1-\gamma}{\gamma}\right) d_n = (n \pi)^2  \left(\frac{1-\gamma}{\gamma}\right) \sum_{m=1}^{\infty}  d_m \sigma_{mn} - \frac{\phi}{\epsilon \gamma}  f_n,
\end{equation}
\begin{equation}
(n\pi)^2 (1-\alpha) \left(\frac{\gamma}{1 - \gamma}\right) c_n+ \left( (n \pi)^2 - s_2 \right) d_n = (n \pi)^2 \sum_{m=1}^{\infty}  d_m \sigma_{mn}
- \frac{\phi}{\epsilon(1 - \gamma)} f_n.
\end{equation}
Solving this system for the coefficients $c_n$ and $d_n$ gives
\begin{equation}
c_n = \frac{s_2}{D_n} \left( \frac{1 - \gamma}{\gamma} \right) \left[ \frac{\phi}{\epsilon (1 - \gamma)} f_n - (n \pi)^2 \sum_{m = 1}^{\infty} d_m \sigma_{mn} \right], \label{An}
\end{equation}
\begin{equation}
d_n = \frac{s_1 - \alpha (n \pi)^2 }{D_n} \left[ \frac{\phi}{\epsilon (1 - \gamma)} f_n - (n \pi)^2 \sum_{m = 1}^{\infty} d_m \sigma_{mn} \right], \label{Bn}
\end{equation}
where
\begin{equation}
D_n = s_1 s_2 - \left( s_1 + s_2 \right) (n \pi)^2 + \alpha (n \pi)^4.
\end{equation}
Note that we have written the bottom topography as an extra ``forcing" term. Indeed, equation \eqref{Bn} for $d_n$ may be viewed as an infinite system of linear equations. To highlight this fact, it can be written as $\textbf{Md} = \boldsymbol{f}$, where 
\begin{equation}
\textbf{M}_{nm} = \begin{cases} 
D_n + (s_1- \alpha (n \pi)^2) (n \pi)^2 \sigma_{nm} & \text{for } n = m\\
 (s_1- \alpha (n \pi)^2) (n \pi)^2 \sigma_{nm} & \text{for } n \neq m
\end{cases},
\end{equation}
and
\begin{equation}
\boldsymbol{f}_{n} =\left(s_1 -\alpha(n  \pi)^2\right)  \frac{\phi}{\epsilon (1 - \gamma)} f_n.
\end{equation}
For a finite number of modes this system can be solved directly for $\mathbf{d} = (d_1, d_2, \dots)$ by simply inverting the matrix $\textbf{M}$. From this we may solve for the $c_n$ and using equations \eqref{mass11}, \eqref{mass22} we can calculate the elevations
\begin{equation}
N_1(x) =\sum_{n=1}^{\infty} a_n \cos n\pi x, \quad 
N_2(x) =\sum_{n=1}^{\infty} b_n \cos n\pi x, \label{Fouriersol}
\end{equation}
with
\begin{equation}
a_n = i\epsilon n \pi\left(\gamma c_n + (1-\gamma)\left(d_n-\sum_{m=1}^{\infty}  d_m \sigma_{mn}\right)\right),
\end{equation}
\begin{equation}
b_n = i\epsilon n \pi(1-\gamma)\left(d_n-\sum_{m=1}^{\infty}  d_m \sigma_{mn}\right).
\end{equation}

%%%%%%%%%

\subsection{Neumann series solution}

For small enough topography, we can obtain a solution for both finite or infinite basins using a Green's function approach. This method is thus limited by the vertical scale of the topography, but provides a natural generalization to the infinite basin and gives a natural decomposition of the system into barotropic and baroclinic scales; this is discussed further in the next subsection. For the rest of this subsection we assume $\alpha \neq 0$, ensuring that $\textbf{B}$ is non-singular, and $\delta_1 \neq \delta_2$. Technical details aside, solutions for the one-layer case may be found by taking the limit $\alpha \to 0$.

We may nearly uncouple the system \eqref{Umatsystem} by diagonalizing the matrix $\textbf{B}$; $\textbf{B} \textbf{V} = \textbf{V} \boldsymbol{\Lambda}$, where $\boldsymbol{\Lambda}$ is a diagonal matrix of the eigenvalues of $\textbf{B}$, in decreasing magnitude, and $\textbf{V}$ is a matrix with the respective eigenvectors. This allows us to change basis, left-multiplying the system by $\textbf{V}^{-1}$, and rewrite \eqref{Umatsystem} as
\begin{equation}
\textbf{A} \textbf{W}(x) + \boldsymbol{\Lambda} \textbf{W}''(x) = \textbf{V}^{-1} \textbf{F}(x) -  \sigma \textbf{A} \textbf{V}^{-1} \textbf{e}_{22} \left( \frac{b(x)}{1 - \sigma b(x)} \right) \textbf{V} \textbf{W}(x), \label{Wmatsystem}
\end{equation}
which fully uncouples the left-hand side terms of \eqref{Wmatsystem}. This is easily seen using the fact that $\textbf{A}$ commutes with any other matrix. The new variable $\textbf{W}(x) = \left[\begin{smallmatrix} U^{BT}(x)\\ U^{BC}(x)\end{smallmatrix} \right]:= \textbf{V}^{-1} \textbf{U}(x)$, which represents a separation of the system into large and small scale motions, can be thought of as representing the barotropic ($BT$) and baroclinic ($BC$) modes of the system. In this case, we associate the first term $U^{BT}(x)$ with the motions arising from the smaller of the eigenvalues (larger scales), we call this the barotropic mode, and the second term $U^{BC}(x)$ with the motions arising from the larger of the two eigenvalues (smaller scales), we call this the baroclinic mode. The reader should be warned that our definition of the separation may not be consistent with other definitions of the words barotropic and baroclinic in the literature.

We rewrite \eqref{Wmatsystem} as
\begin{equation}
\textbf{E} \textbf{W}(x) +\textbf{W}''(x) = \boldsymbol{\Lambda}^{-1} \textbf{V}^{-1} \textbf{F}(x) -  \sigma \textbf{E} \textbf{V}^{-1} \textbf{e}_{22} \left( \frac{b(x)}{1 - \sigma b(x)} \right) \textbf{V} \textbf{W}(x), \label{Wsolvesystem}
\end{equation}
where $\textbf{E} := \boldsymbol{\Lambda}^{-1} \textbf{A} = \left[\begin{smallmatrix} (k^{BT})^2 & 0\\ 0 & (k^{BC})^2 \end{smallmatrix} \right]$. Note that $k^{BT}$ and $k^{BC}$ arise from the eigenvalues of $\textbf{B}$ and satisfy the polynomial equation
\begin{equation}
k^4 - \left( \frac{s_1 + s_2}{\alpha} \right) k^2+ \frac{s_1 s_2}{\alpha} =0. \label{poly}
\end{equation}
In the case $\alpha \ll 1$, typical for the ocean, it follows that $|k^{BT}| \ll |k^{BC}|$. In other words, $k^{BT}$ and $k^{BC}$ define two well separated scales for the problem. To avoid ambiguity about which solution of \eqref{poly} we refer to, we pick those solutions such that $\text{Im}(k^{BT}), \text{Im}(k^{BC}) >0$. That is, we label $k^{BT}$ and $k^{BC}$ the solutions of \eqref{poly} in the upper half of the complex plane.

In the low damping regime  ($\delta_1, \delta_2 \ll 1$), the polynomial \eqref{poly} can be written in the approximate form
\begin{equation}
k^4 - \left( \frac{s_1 + s_2}{\alpha} \right) k^2 + \frac{s_1 s_2}{\alpha} \approx k^4 - \frac{k^2}{\epsilon^2 \alpha \gamma  (1-\gamma)} + \frac{1}{\epsilon^4 \alpha \gamma (1-\gamma)} =0.
\end{equation}
Writing the equation in this manner clearly shows that in the low damping regime we essentially have one stratification parameter: $\alpha \gamma (1-\gamma)$. For this reason, perturbations in $\alpha$ may lead to the same scales $k^{BT}$ and $k^{BC}$ as perturbations in $\gamma$. In particular, in the case $\gamma < 1/2$ which represents a thin upper layer over a thicker lower layer, a positive perturbation in $\alpha$ (increased density contrast) may lead to equivalent scales as a positive perturbation in $\gamma$ (increased upper layer thickness). Lastly, it is worth noting that in this low damping regime equation \eqref{poly} is equivalent to the formula for the phase speed of two superposed fluids with different density;
\begin{equation}
c^4 - g (H_1 + H_2) c^2 + g g' H_1 H_2 =0
\end{equation}
as given by \cite{Gill1982} under the substitution $c = \omega L/k$.

The new system \eqref{Wsolvesystem} can be readily solved using the Green's function of the problem
\begin{equation}
\textbf{E} \textbf{W}(x) + \textbf{W}''(x) = \textbf{f}(x), \label{greenssystem}
\end{equation}
where $\textbf{f}(x)$ stands for the right-hand side of \eqref{Wsolvesystem}. The solution of \eqref{greenssystem} is given by
\begin{equation}
\textbf{W}(x) = \int_{\ell_1}^{\ell_2} \textbf{G}(x,y) \textbf{f}(y)\; dy \quad \text{ with } \quad
\textbf{G}(x,y) = \begin{bmatrix} g^{BT}(x,y) & 0\\ 0 & g^{BC}(x,y) \end{bmatrix}, \label{WGsolve}
\end{equation}
where $\textbf{G}(x,y)$, $\ell_1 < x,y < \ell_2$, satisfies the system
\begin{equation}
\textbf{E} \textbf{G}(x,y) + \textbf{G}_{yy}(x,y)= \delta(y - x) \mathbb{I} \label{greenseqn}
\end{equation}
with boundary conditions
\begin{equation}
\textbf{G}(x,\ell_1) = \textbf{G}(x,\ell_2)=\textbf{0}.
\end{equation}
Note that $g^{BT}(x,y)$ is the Green's function arising from the barotropic mode, top equation of system \eqref{greenssystem}, and $g^{BC}(x,y)$ is the Green's function arising from the baroclinic mode, bottom equation of system \eqref{greenssystem}. Moreover, all information about the boundary conditions is contained in the Green's functions. This implies that we require different Green's functions for the finite and infinite basin cases. The Green's functions are
\begin{equation}
g(x,y) = \begin{cases} \frac{\sin k (x - \ell_1) \sin k (y - \ell_2)}{k \sin k(\ell_2 - \ell_1)} & x < y\\
 \frac{\sin k (y - \ell_1) \sin k (x - \ell_2)}{k \sin k(\ell_2 - \ell_1)} & y < x \end{cases} \quad \text{ and } \quad g(x,y) = \begin{cases} \frac{e^{i k (y-x)}}{2 i k} & x < y\\
 \frac{e^{ik (x-y)}}{2 i k} & y < x \end{cases} \label{greensdef}
\end{equation}
for a finite ($\ell_1 < \ell_2$) and infinite basin, respectively; see for example \cite{Greenberg1971}. In this case, $k$ is a placeholder for $k^{BT}$ or $k^{BC}$ depending on whether $g(x,y)$ is meant to denote $g^{BT}(x,y)$ or $g^{BC}(x,y)$, respectively. Moreover, the construction of the Green's function for the infinite basin follows from the fact that we impose $\text{Im}(k) >0$, as previously stated.

Now, with the adequate Green's functions for the basin in question, the solution method \eqref{WGsolve} allows us to write
\begin{equation}
\textbf{W}(x) = \int_{\ell_1}^{\ell_2} \textbf{D}(x,y) \textbf{F}(y) \; dy- \sigma s_1 s_2 \int_{\ell_1}^{\ell_2} \textbf{D}(x,y) \textbf{e}_{22} \left( \frac{b(y)}{1 - \sigma b(y)} \right) \textbf{V} \textbf{W}(y) \; dy, \label{Weqns}
\end{equation}
where $\textbf{D}(x,y) = \textbf{G}(x,y) \boldsymbol{\Lambda}^{-1} \textbf{V}^{-1}$. Letting
\begin{equation}
 \textbf{F}_0(x) :=  \int_{\ell_1}^{\ell_2} \textbf{D}(x,y) \textbf{F}(y) \; dy
\end{equation}
and defining an operator $K$ by
\begin{equation}
(K \textbf{W})(x) := - s_1 s_2 \int_{\ell_1}^{\ell_2} \textbf{D}(x,y) \textbf{e}_{22} \left( \frac{b(y)}{1 - \sigma b(y)} \right) \textbf{V} \textbf{W} (y) \; dy \label{KU}
\end{equation}
we can concisely write the equation \eqref{Weqns} as $\textbf{W}(x) = \textbf{F}_0(x)+\sigma K \textbf{W}(x)$. This last expression may be solved by the formal Neumann series
\begin{equation}
\textbf{W}(x) = \textbf{F}_0(x) + \sigma K \textbf{F}_0(x) + \sigma^2 K^2 \textbf{F}_0(x) + \cdots \label{geomsol}
\end{equation}
giving the full solution of $\textbf{U}(x) = \textbf{V} \textbf{W}(x)$ in either a finite or infinite basin when the topography $\sigma$ is small enough that the series \eqref{geomsol} converges.

%%%% Separation into baro modes %%%%%%%%

\subsection{Separation into barotropic and baroclinic modes}\label{sec:separa}

A separation for the solution is suggested in the previous subsection.
To make multiplication by $\textbf{V}$ completely determined we normalize the eigenvectors as presented in \cite{Flierl1978} and \cite{Muller2006}. Essentially, for a given eigenvector $\textbf{v}=(v_1,v_2)$ we impose the condition $\gamma v_1^2+(1-\gamma) v_2^2 = 1$; a depth weighted normalization. Under these assumptions we may write the matrix $\textbf{V}$ of eigenvectors as
\begin{equation}
\textbf{V} = \begin{bmatrix} c^{BT}(1-\gamma) & c^{BC} (1-\gamma) \\ c^{BT} \gamma \left( \frac{s_1}{(k^{BT})^2} - 1\right) & c^{BC} \gamma \left( \frac{s_1}{(k^{BC})^2}  - 1\right) \end{bmatrix},\label{Veqn}
\end{equation}
where
\begin{equation}
c^{BT,BC} = \frac{1}{\sqrt{(1-\gamma) \gamma \left(  1-\gamma + \gamma \left( \frac{s_1}{(k^{BT,BC})^2} - 1\right)^2 \right)}  }
\end{equation}
to satisfy the normalization conditions, and
\begin{equation}
\textbf{V}^{-1} = \frac{1}{s_1} \left( \frac{1}{(k^{BT})^2} - \frac{1}{(k^{BC})^2} \right)^{-1} \begin{bmatrix} 
-\frac{1}{c^{BT}(1-\gamma)} \left( \frac{s_1}{(k^{BC})^2} - 1\right) & \frac{1}{c^{BT}\gamma}\\ \frac{1}{c^{BC} (1-\gamma)} \left( \frac{s_1}{(k^{BT})^2} - 1\right) & -\frac{1}{c^{BC} \gamma}
\end{bmatrix}.
\end{equation}
Thus, the matrix $\textbf{V}$ is invertible provided that $(k^{BT})^2 \neq (k^{BC})^2$. In the case when $\alpha$ is small the scales of the problem are well separated making $\textbf{V}$ invertible. We use the notation $c^{BT,BC}$ to mean that $c^{BT}$ can be obtained by only using the first superscript in every term in the equation and $c^{BC}$ by only using the second superscript. For example, in the case of $c^{BT}$ we only use $k^{BT}$ and for $c^{BC}$ we only use $k^{BC}$. It may not be the case that the first superscript always refers to the barotropic mode; see the first term of equation \eqref{UBTBC} below.

For the layer elevations, a separation can be achieved simply by rewriting \eqref{Neq} as $\textbf{N}(x) =i \epsilon \left( \mathbb{I} + \textbf{e}_{12} \right) \boldsymbol{\Gamma} \textbf{V}\textbf{W}'(x)$. In this manner we can define the barotropic and baroclinic components of the surface and interfacial elevations as those arising from $U^{BT}$ and $U^{BC}$, respectively.
From matrix multiplication it follows that
\begin{equation}
N_1^{BT,BC} = i \epsilon (1-\gamma) \gamma c^{BT,BC} \left( \frac{s_1}{(k^{BT,BC})^2}  \right) \frac{d U^{BT,BC}}{dx}(x), \label{N1BTBC}
\end{equation}
\begin{equation}
N_2^{BT,BC} = i \epsilon (1-\gamma) \gamma c^{BT,BC} \left( \frac{s_1}{(k^{BT,BC})^2} - 1\right) \frac{d U^{BT,BC}}{dx}(x).
\end{equation}
The equations for $U^{BT}$ and $U^{BC}$, equation \eqref{Weqns}, can similarly be written as
\begin{equation}
\begin{split}
U^{BT,BC} =&   \frac{1}{s_1 c^{BT,BC}} \left( \frac{1}{(k^{BT,BC})^2} - \frac{1}{(k^{BC,BT})^2} \right)^{-1} \left[ \frac{\phi}{\gamma (1-\gamma) \epsilon} \int_{\ell_1}^{\ell_2} g^{BT,BC} e^{i \phi y} \; dy  \right. \\
  & + \sigma (k^{BT,BC})^2 c^{BT} \left( 1 - \frac{s_1}{(k^{BT})^2} \right) \int_{\ell_1}^{\ell_2}  g^{BT,BC} U^{BT} \left( \frac{b}{1 - \sigma b}\right)\; dy \\
  & +\left. \sigma (k^{BT,BC})^2  c^{BC} \left( 1 -  \frac{s_1}{(k^{BC})^2} \right) \int_{\ell_1}^{\ell_2}g^{BT,BC} U^{BC} \left( \frac{b}{1 - \sigma b}\right) \; dy \right].
\end{split}\label{UBTBC}
\end{equation}
We have written $g^{BT}$ as opposed to $g^{BT}(x,y)$, $b$ as opposed to $b(y)$, etc., to avoid clutter. It should be clear from context which terms are functions and which are parameters.

Equation \eqref{UBTBC} clearly shows that in the presence of topography there is mixing between the barotropic and baroclinic modes. Note that the barotropic solution $U^{BT}$ depends in a non-trivial manner on the solution $U^{BC}$ and that this coupling occurs only when the topography is non-zero, as shown in the last term of \eqref{UBTBC}. On the other hand only the barotropic Green's function appears in the barotropic solution and only the baroclinic Green's function appears in the baroclinic solution. This corroborates that our splitting into barotropic and baroclinic modes is reasonable. 

%%%%%%%%%%%

\subsection{Scattering solution}

A perturbation expansion with respect to the topographical parameter $\sigma$ in equations \eqref{mass1}$-$\eqref{mom2} shows that the equations obtained at order $\sigma$ for the infinite basin give the classical topographical scattering solution in terms of an incident velocity. That is, the $\sigma$ order terms of the solution correspond to the scattering solution of the barotropic tide on the bottom roughness. The problem of internal wave generation from a bump that is impinged upon by a barotropic tidal velocity has received much attention in the oceanographic community; see for example \cite{Llewellyn+al2001}, \cite{StLaurent+al2003}, \cite{Khatiwala2003}. We briefly derive the scattering solution in this subsection to highlight its similarities with our full solution.

Consider expanding all variables in the problem in terms of the topographical parameter $\sigma$, i.e., $\varphi(x) \sim \varphi_0(x) + \sigma \varphi_1(x) + \cdots$ for $\sigma \to 0$, where $\varphi(x)$ is the variable of interest. Gathering the order $\sigma$ terms in equations \eqref{mass1}$-$\eqref{mom2}, after using the separation \eqref{separation}, gives
\begin{equation}
i \left( N_1 - N_2 \right) + \epsilon \gamma U_1' = 0,
\end{equation}
\begin{equation}
 i N_2 + \epsilon (1 - \gamma) U_2' = \epsilon (1 - \gamma)F,
\end{equation}
\begin{equation}
i U_1 + \epsilon N_1' + \delta_1 U_1=0,
\end{equation}
\begin{equation}
i U_2 +(1-\alpha) \epsilon N_1' + \alpha \epsilon N_2' + \delta_2 U_2 = 0,
\end{equation}
where $F = \frac{\partial}{\partial x}\left( U_{2,0} b \right)$ with $U_{2,0}$ representing the incident tidal velocity on the topography or equivalently the $O(1)$ term in the expansion of $U_2$. For the scattering problem, we assume that this background state $U_{2,0}$ is known and focus on solving for the higher order terms. Note that all variables in this equation correspond to the order $\sigma$ terms, but we have dropped any indicative subscripts to avoid cumbersome notation. That is, in this subsection the notation $N_1$ really stands for $N_{1,1}$, the order $\sigma$ term in the expansion of $N_1$.

Since the system obtained is linear and has constant coefficients, we may readily solve the system above in the infinite basin case using the Fourier transform, e.g., $\hat{u}(k) = \int_{-\infty}^{\infty} u(x) e^{-i kx} \; dx$. Manipulating the equations this yields
\begin{equation}
\left(s_1 - k^2 \right) \hat{N}_1 = s_1 \hat{N}_2,
\end{equation}
\begin{equation}
\left( s_2 - \alpha k^2 \right) \hat{N}_2 - (1 - \alpha) k^2 \hat{N}_1 = - i \epsilon (1-\gamma) s_2 \hat{F},
\end{equation}
where $s_1$ and $s_2$ are as introduced previously. These equations may be combined to obtain
\begin{equation}
\left( k^4 - \left(\frac{s_1 + s_2 }{\alpha }\right) k^2+ \frac{s_1 s_2}{\alpha } \right) \hat{N}_1 = - i \epsilon (1-\gamma) \frac{s_1 s_2}{\alpha} \hat{F}.\label{N1hat}
\end{equation}
It is worth noting that the left-hand side of this equation is \eqref{poly}, making the separation of scales derived here identical to that in the previous section. This allows us to write equation \eqref{N1hat} as 
\begin{equation}
\hat{N}_1 = - i \epsilon (1-\gamma) (k^{BT})^2 (k^{BC})^2 \left( \frac{1}{(k - k^{BT})(k+k^{BT})(k-k^{BC})(k+k^{BC})} \right) \hat{F} = \hat{R}_1 \hat{F},\label{N1hatsolve}
\end{equation}
where $k^{BT}$ and $k^{BC}$ are the roots of the polynomial \eqref{poly} as defined previously. Therefore, we obtain a closed form solution for $N_1(x)$ -- the order $\sigma$ term in the surface elevation -- by taking the inverse Fourier transform on $\eqref{N1hatsolve}$. The problem of inverting $\hat{N}_1(x)$ in equation \eqref{N1hatsolve} to obtain $N_1(x)$ is of course equivalent to taking a convolution of the functions $R_1(x)$ and $F(x)$. For this reason we seek a closed form solution of $R_1(x)$ since $F(x)$ is the known forcing function.

The integral $R_1(x) = \frac{1}{2\pi}\int_{-\infty}^{\infty} \hat{R}_1(k) e^{ikx}\; dk$ is readily done by summing residues separately for $x \geq 0$ and for $x < 0$ since $k^{BT}$ and $k^{BC}$ lie in the upper complex plane. In the case of $x \geq 0$ we close the contour of integration around the upper half plane leading to a positively oriented curve and in the case $x <0$ we close the contour of integration around the lower half plane leading to a negatively oriented curve.
After some manipulation we find  $R_1(x) = R_1^{BT}(x) + R_1^{BC}(x)$, where
\begin{equation}
R_1^{BT}(x) = -\frac{\epsilon (1-\gamma)}{2} \frac{e^{i k^{BT} |x|}}{k^{BT}} \left(\frac{1}{(k^{BT})^2}-\frac{1}{(k^{BC})^2}\right)^{-1} \label{R1BT}
\end{equation}
and
\begin{equation}
R_1^{BC}(x) = \frac{\epsilon (1-\gamma)}{2} \frac{e^{ik^{BC} |x|}}{k^{BC}} \left(\frac{1}{(k^{BT})^2}-\frac{1}{(k^{BC})^2}\right)^{-1}. \label{R1BC}
\end{equation}
This separation of $R_1$ leads to a very natural way of writing the upper layer solution in terms of its barotropic and baroclinic components. Explicitly, we write
\begin{equation}
N_1^{BT}(x) = \int_{-\infty}^{\infty} R_1^{BT}(x-y) F(y) \; dy \quad \text{ and } \quad N_1^{BC}(x) = \int_{-\infty}^{\infty} R_1^{BC}(x-y) F(y) \; dy \label{eta1t}
\end{equation}
for the order $\sigma$ barotropic and baroclinic components of the surface elevation, respectively. 

Due to the fact that the scattering solution requires prior knowledge of the incident barotropic tide we do not include plots or analysis on it, but rather do so -- in the next section -- for the full solution. 

%%%%%%%%%%%%%%%%

\section{Effects of stratification on the surface elevation}\label{sec:num}

In this section we examine the impacts of stratification on surface tidal elevations in both the finite and infinite basin.
We also examine the importance that topography and basin boundaries have on the value of the perturbed surface elevation.
We utilize a suitably truncated versions of the Fourier series solution \eqref{Fouriersol} and Neumann series solution \eqref{geomsol} for several physically motivated trials. 

\subsection{Parameter values and description of trials} \label{sec:params}

For the finite basin we impose boundaries at $\ell_1 = -1$, $\ell_2 = 0$ and for the infinite basin we assume that both boundaries lie at infinity. For the Neumann series, the only significant computational difference between the two basins lies in the different Green's functions employed; equation \eqref{greensdef}. 
For both basins we utilize an astronomical forcing with the parameter $\phi \approx 1.2827$, arising from $k = \frac{2}{6,371 \times 10^{3}}$ m$^{-1}$ ($2 \pi$ over the zonal wavelength of the semidiurnal tidal potential at the equator) and $L= 4,086 \times 10^{3}$ m (a typical ocean basin scale) as in \cite{Arbic+al2009}; recall $\phi = kL$.

We choose a Gaussian topography of the form
\begin{equation}
b(x) = \frac{H_0}{H_2} e^{-(x+0.5)^2/a^2}
\end{equation}
so that it is centered in the finite basin. The parameter $a$ is chosen as $a = \frac{\pi}{2 a_1 L}$, where $a_1 = \omega \sqrt{(H_1 + H_2)/(g' H_1 H_2)}$ is the topographic wavenumber that excites the first baroclinic mode in the absence of damping.

For all plots, except where noted, we use the forcing frequency $\omega = 1.405189 \times 10^{-4}$ s$^{-1}$ (the $M_2$ frequency), $H_1 = 700$ m, $H_2 = 3300$ m, and a reduced gravity $g'=1.64 \times 10^{-2}$ m s$^{-2}$. As an aside, even though in the different trials the values of $g'$, $H_1$, and $H_2$ change, we keep the value of $a$ -- obtained using $g'$, $H_1$, and $H_2$ above -- constant throughout. Other parameters used will be as follows: $g = 9.81$ m s$^{-2}$, and $\delta_2 \approx 0.0412$ as the canonical value for the open ocean used in \cite{Arbic+al2009}. Thus, unless mentioned otherwise, we use the following values for all plots: $\alpha \approx 0.0017$, $\epsilon \approx 0.345$, and $\gamma = 0.175$. In addition, we set $H_0 = 2,350$ m and $\sigma = 1$. The value of $H_0$ was chosen so that the ratio of the spatial and temporal averaged square of the upper layer and lower layer velocity in our finite basin control case (ratio $\approx 2.5$) is approximately equal to the value obtained for these parameters in the global $1/8^{\circ}$ simulations discussed in section \ref{sec:numerical}.

For each basin type we consider four trials: one layer solution, two layer solution (referred to as control) with $\alpha \approx 0.0017$, two layer solution with $\alpha \approx 0.0018$, and two layer solution with $\alpha \approx 0.0017$ -- as in the control -- but with $\gamma =0.2$. In this way, the trials cover the cases of unstratified flow, stratified flow, perturbation of the layers' densities, and perturbation in the layers' thicknesses; as in Figures \ref{fig:global_AMP} and \ref{fig:global_PHASE}. We obtain $\alpha \approx 0.0018$ by assigning $g' = 1.78 \times 10^{-2}$ m s$^{-2}$ and $\gamma = 0.2$ by assigning $H_1 = 800$ m. These values of $\alpha$ and $\gamma$ are intended to represent climatic perturbations to the values of $g'$ and $H_1$.  The interface perturbation is motivated by Figure 10 of \cite{Arbic+Owens2001}, which shows $\sim$100 m displacements of isopycnals over decadal timescales in hydrographic observations of the North Atlantic. The $g'$ perturbation was estimated from the $0.5^{\circ}$C century$^{-1}$ nominal maximum warming trend found in intermediate depth waters in the same paper.  We computed the change in $g'$ (with potential densities referenced to  1780 db) that would take place if a water parcel at 100 db having salinity 37 psu and temperature of $20^{\circ}$C warmed by $0.5^{\circ}$C.  The deep reference parcel has depth of $3000$ db, salinity $34.5$ psu and temperature of $4^{\circ}$C.

\begin{figure}[h!]
\begin{center}
(a) One layer solution, $\alpha = 0$\\
\includegraphics[scale=0.65]{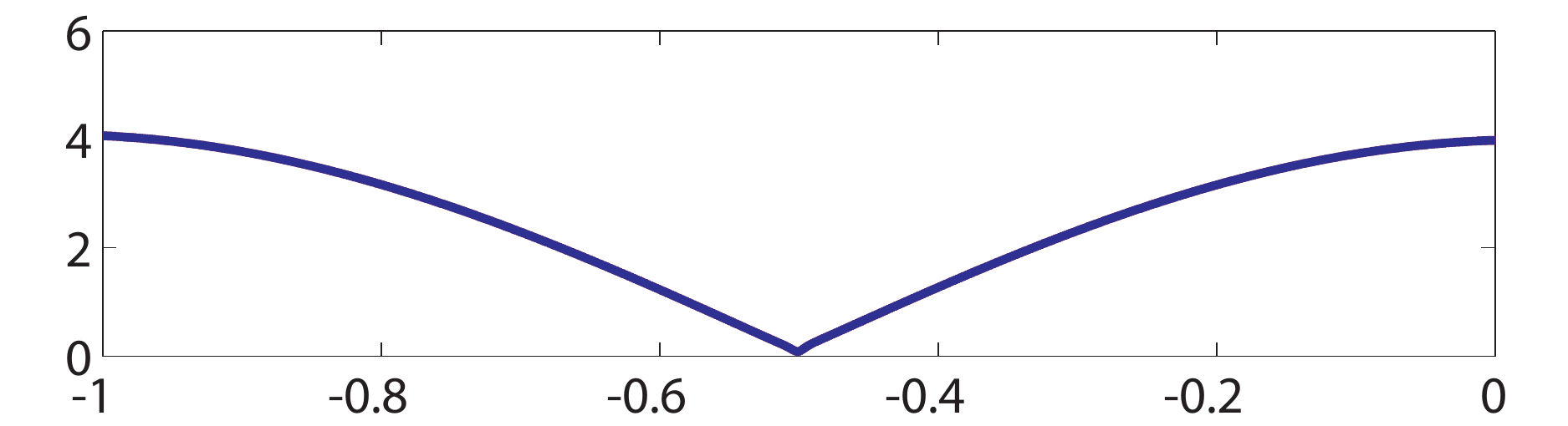} $x$\\
\vspace{-0.25cm}
(b) Two layer control, $\alpha \approx 0.0017$, minus one layer solution\\
\includegraphics[scale=0.65]{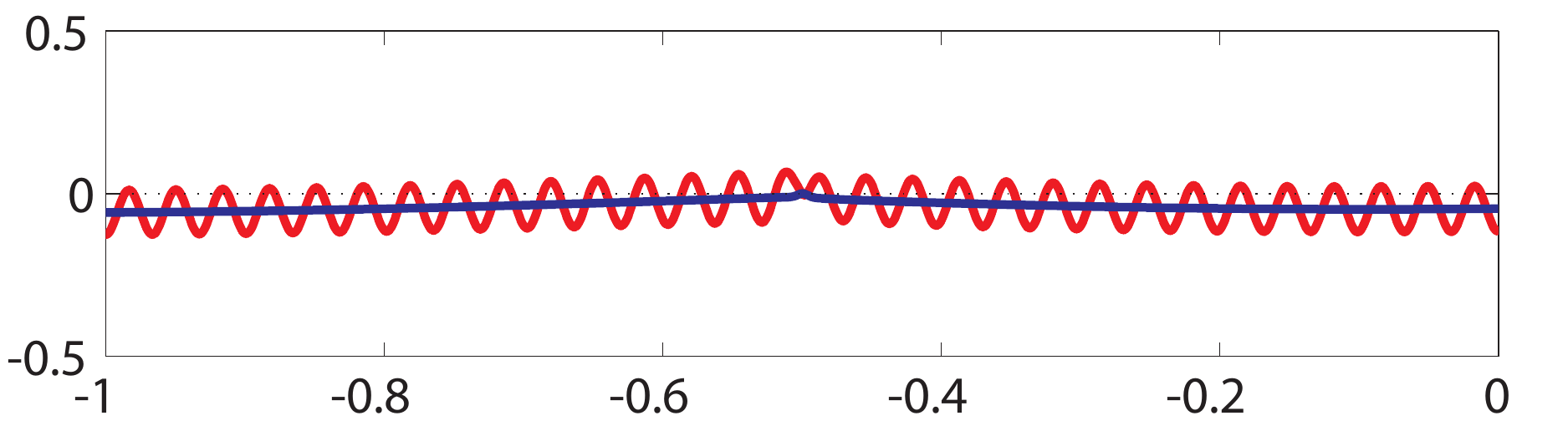} $x$\\
\vspace{-0.25cm}
(c) Two layer, $\alpha \approx 0.0018$, minus two layer control solution\\
\includegraphics[scale=0.65]{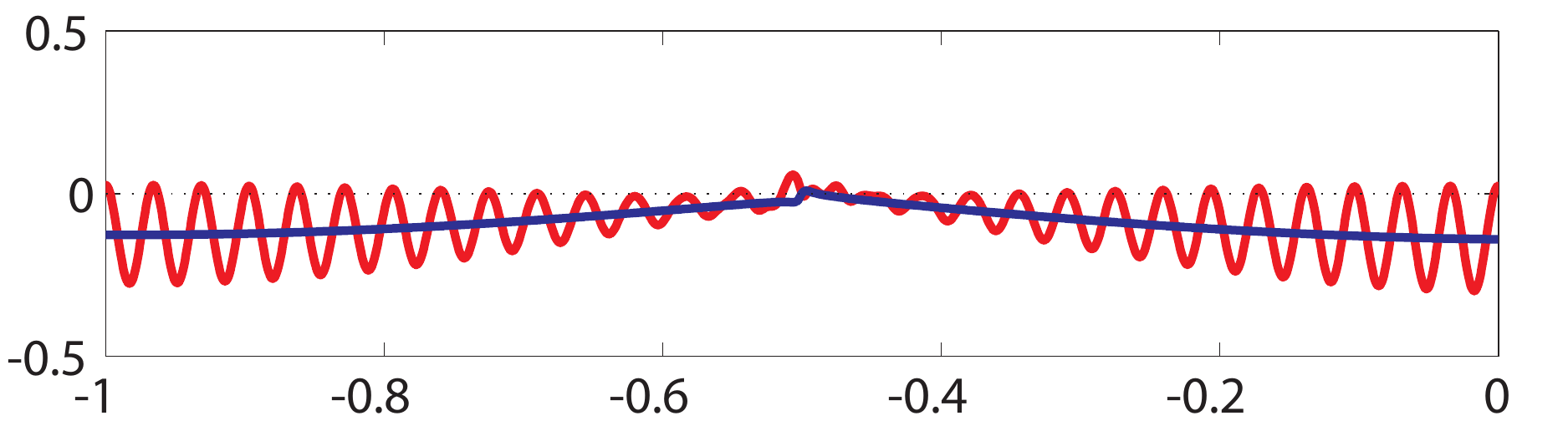} $x$\\
\vspace{-0.25cm}
(d) Two layer, $\gamma = 0.2$, minus two layer control solution\\
\includegraphics[scale=0.65]{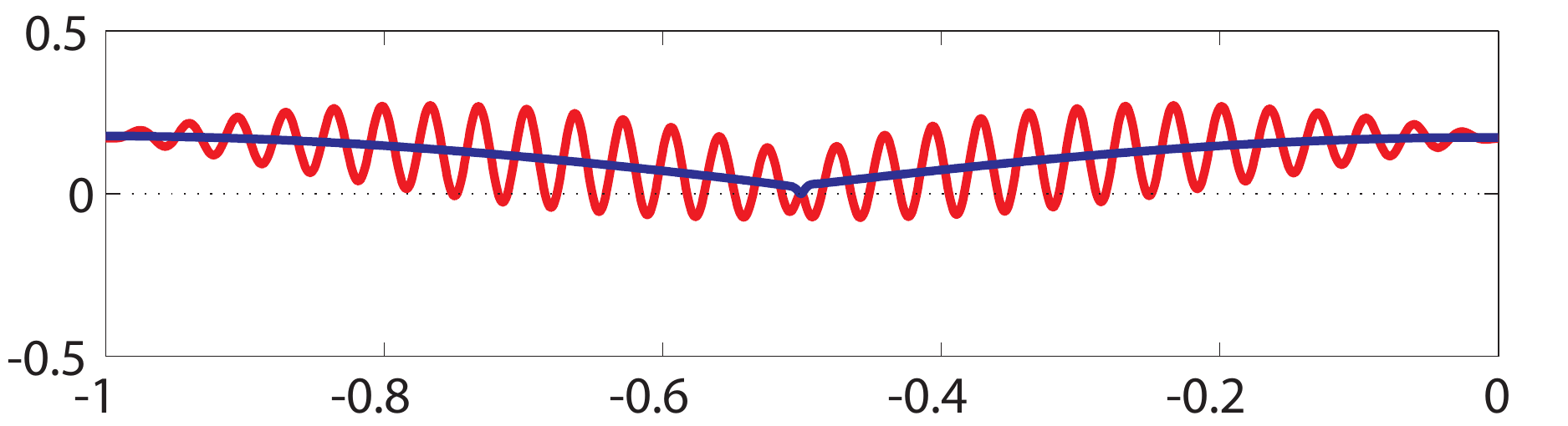}
$x$ \vspace{-0.5cm}
\caption{Surface elevation (red line) and barotropic surface elevation (blue line) amplitudes in a finite basin. Plot (a) was obtained using the Fourier series method while (b), (c), and (d) were obtained using the Neumann series method. The $y$-axis in all plots is non-dimensional amplitude. The range in the vertical scales of (b), (c), and (d) is identical. } \label{fig:fin_magBT}
\end{center}
\end{figure}

\begin{figure}[h!]
\begin{center}
(a) One layer solution, $\alpha = 0$\\
\includegraphics[scale=0.65]{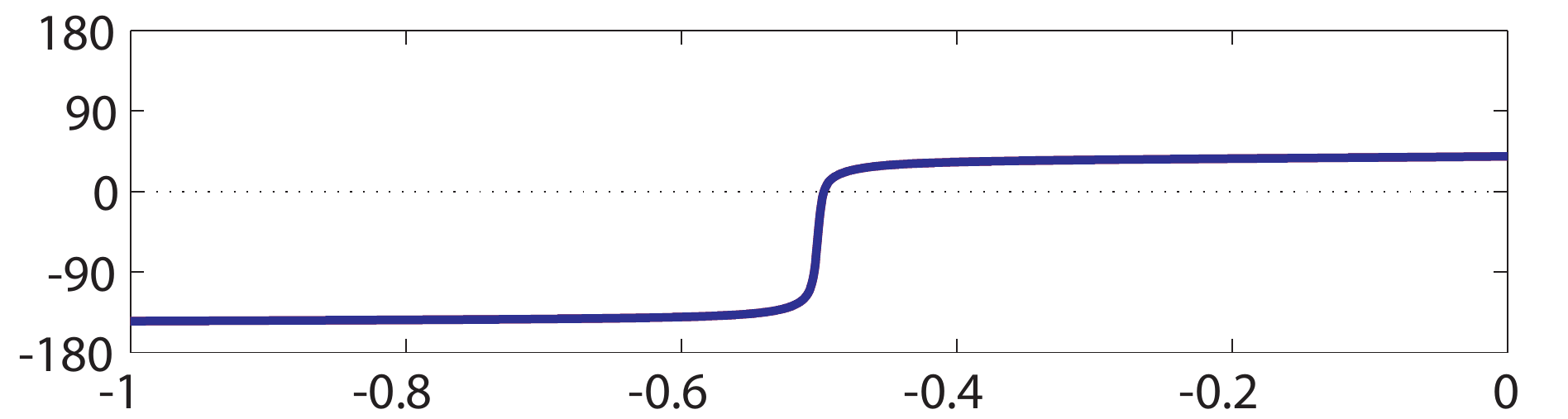} $x$\\
\vspace{-0.25cm}
(b) Two layer control, $\alpha \approx 0.0017$, minus one layer solution\\
\includegraphics[scale=0.65]{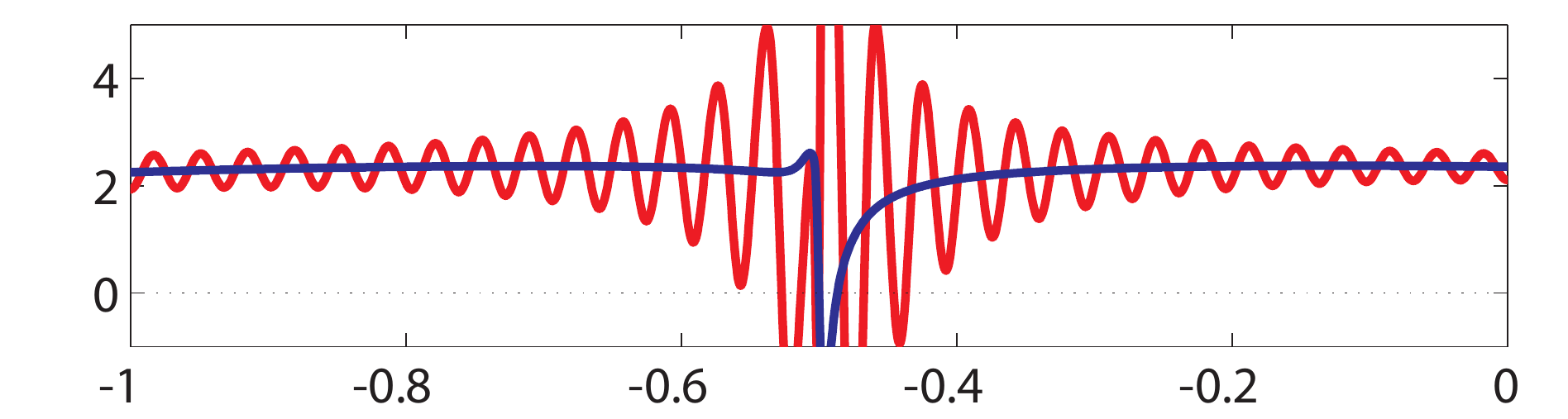} $x$\\
\vspace{-0.25cm}
(c) Two layer, $\alpha \approx 0.0018$, minus two layer control solution\\
\includegraphics[scale=0.65]{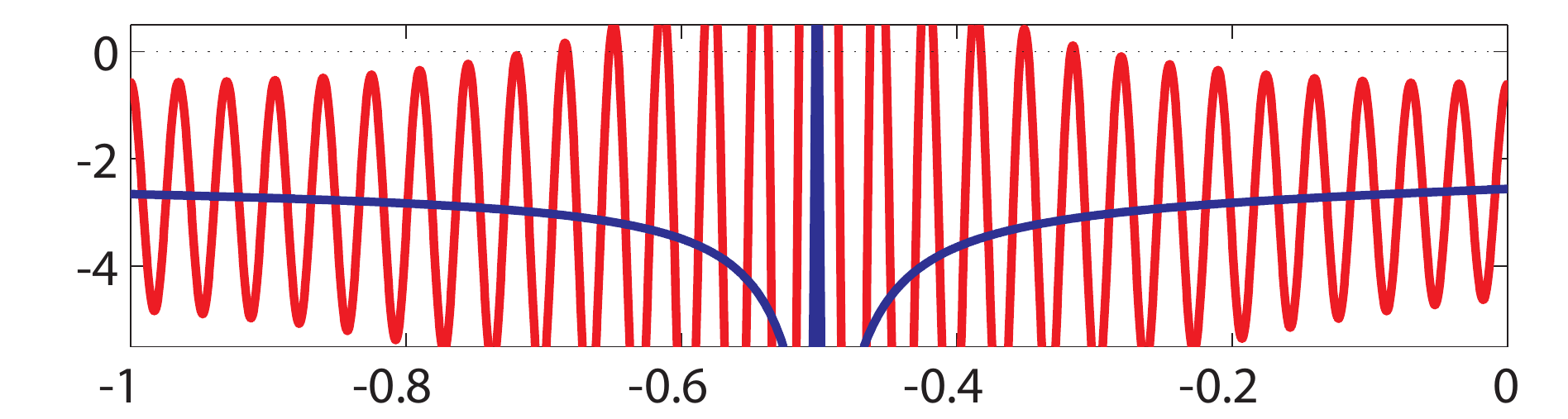} $x$\\
\vspace{-0.25cm}
(d) Two layer, $\gamma = 0.2$, minus two layer control solution\\
\includegraphics[scale=0.65]{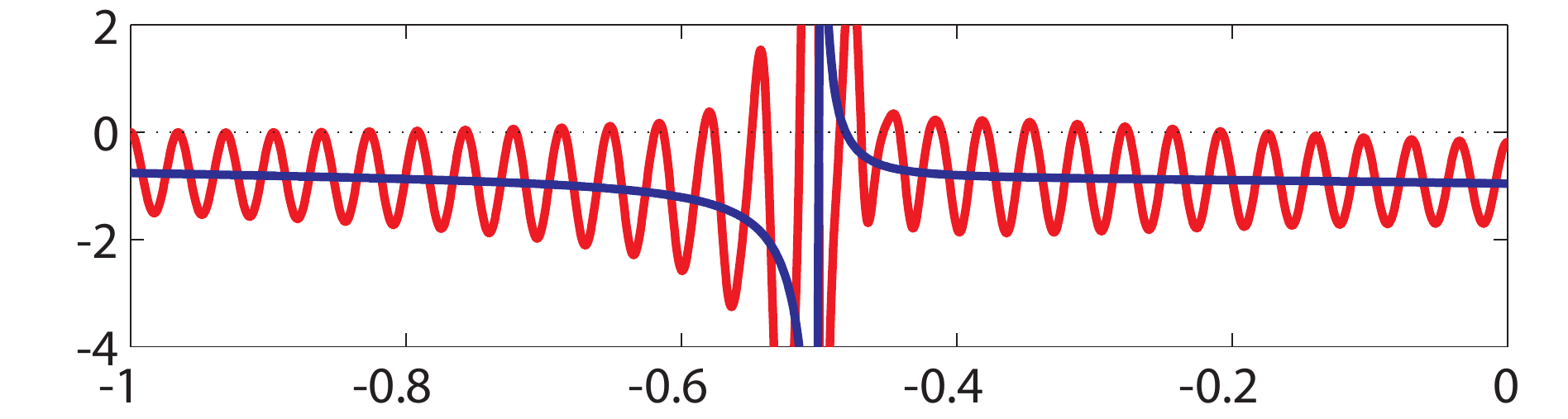}
$x$ \vspace{-0.5cm}
\caption{As in Figure \ref{fig:fin_magBT}, but for phase differences in degrees. 
%All plots centered about the barotropic mode. 
The large phase differences in the center of the basin are not very meaningful since these are associated with small amplitudes; see Figure \ref{fig:fin_magBT} (a). The range in the vertical scales of (b), (c), and (d) is identical.} \label{fig:fin_phaBT}
\end{center}
\end{figure}

The surface elevation amplitudes for the finite basin are presented in Figure \ref{fig:fin_magBT}, and the phases are presented in Figure \ref{fig:fin_phaBT}. Similarly, the amplitudes for the infinite basin are presented in Figure \ref{fig:inf_magBT}, and the phases in Figure \ref{fig:inf_phaBT}. For all figures the surface elevation is given by a solid red line and the barotropic component of the surface elevation is given by a solid blue line. We next explain the setup of figures for the finite basin; the setup in the infinite basin is the same. Figures \ref{fig:fin_magBT} (a) and \ref{fig:fin_phaBT} (a) show the one layer solution.
Figures \ref{fig:fin_magBT} (b) and \ref{fig:fin_phaBT} (b) show the difference between the two layer control, $\alpha \approx 0.0017$, and the one layer solution. Figures \ref{fig:fin_magBT} (c) and \ref{fig:fin_phaBT} (c) show the difference between a two layer solution with a climatically perturbed $\alpha \approx 0.0018$ and the control two layer solution. Lastly, Figures \ref{fig:fin_magBT} (d) and \ref{fig:fin_phaBT} (d) show the difference between a two layer solution with a climatically perturbed $\gamma = 0.2$ and the two layer control. The plots in Figure \ref{fig:fin_percent} are as in Figure \ref{fig:fin_magBT} (b), (c), and (d) but with changes represented as percentages of the barotropic  mode solution of the one-layer control case.

Lastly, we briefly bring up a point which will not be pursued further. For the purposes of this paper, all solutions were found using the Neumann series (Green's function) method except for the finite basin one-layer solution, RMS plots in Figure \ref{fig:ene} and the resonance plot in Figure \ref{fig:resonan} where, instead, the Fourier series expansion was used. 
Figures \ref{fig:ene} and \ref{fig:resonan} will be described in detail in the next section.
For the topography used, the Fourier series solution was more numerically reliable.
The numerical accuracy of all solutions was confirmed by checking that they satisfy the original set of differential equations.

%%%%%%%%%%%

\subsection{Discussion}

Consistent with the numerical results shown in Figures \ref{fig:global_AMP} and \ref{fig:global_PHASE}, plots of the full solution on a finite domain basin show that changes to the stratification lead to changes in the surface elevation's amplitude of order  $2-10 \%$, see Figures \ref{fig:fin_magBT} and \ref{fig:fin_percent}, while changes in the phase of the solution are of order $2-3$ degrees, see Figure \ref{fig:fin_phaBT}.
The addition of a second layer to the finite basin model yields an observable change in the barotropic surface tidal elevation amplitude of up to $1.5\%$ compared with the one layer case; see Figures \ref{fig:fin_magBT} (b) and \ref{fig:fin_percent} (a). A climatic perturbation in $\alpha$ to the two layer solution may induce a further change in the barotropic mode's amplitude of about  $3\%$; see Figures \ref{fig:fin_magBT} (c) and \ref{fig:fin_percent} (b). Climatic perturbations in $\gamma$ may induce qualitatively similar responses from the system; see Figures \ref{fig:fin_magBT} (d) and \ref{fig:fin_percent} (c). Perturbations in $\alpha$ or $\gamma$ may lead to an increase or a decrease in the barotropic mode's amplitude comparable in magnitude to the amplitude perturbation of the baroclinic mode. From inspection of solutions with different topographic amplitudes (not shown here for the sake of brevity), we find that stratification perturbations can yield either an increase or decrease in the barotropic mode, depending sensitively on the topography.

\begin{figure}[h!]
\begin{center}
(a) One layer solution, $\alpha = 0$\\
\includegraphics[scale=0.65]{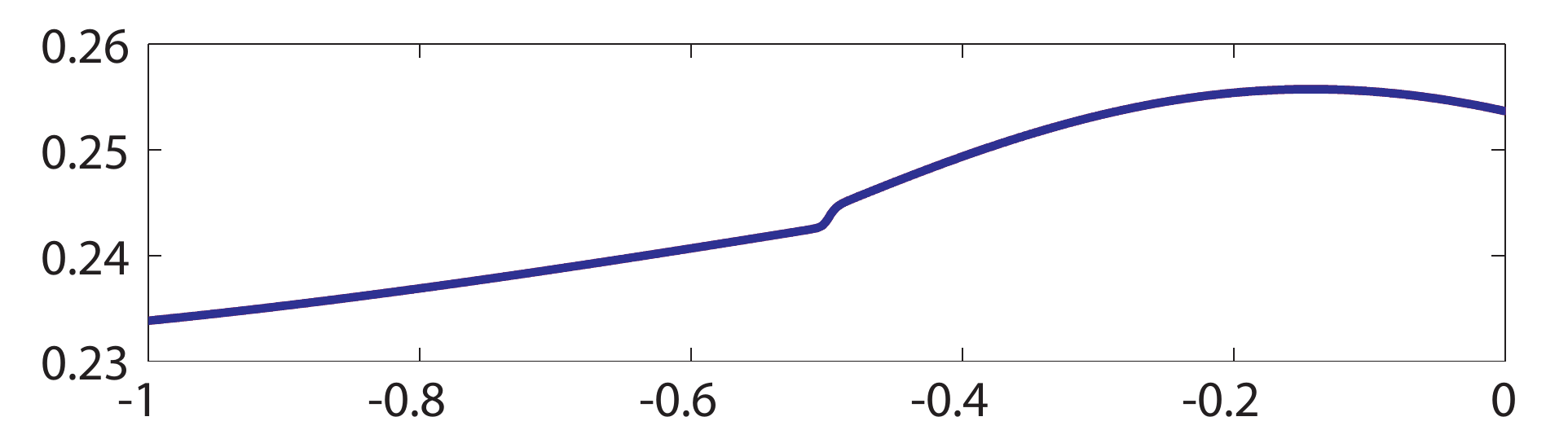} $x$\\
\vspace{-0.25cm}
(b) Two layer control, $\alpha \approx 0.0017$, minus one layer solution\\
\includegraphics[scale=0.65]{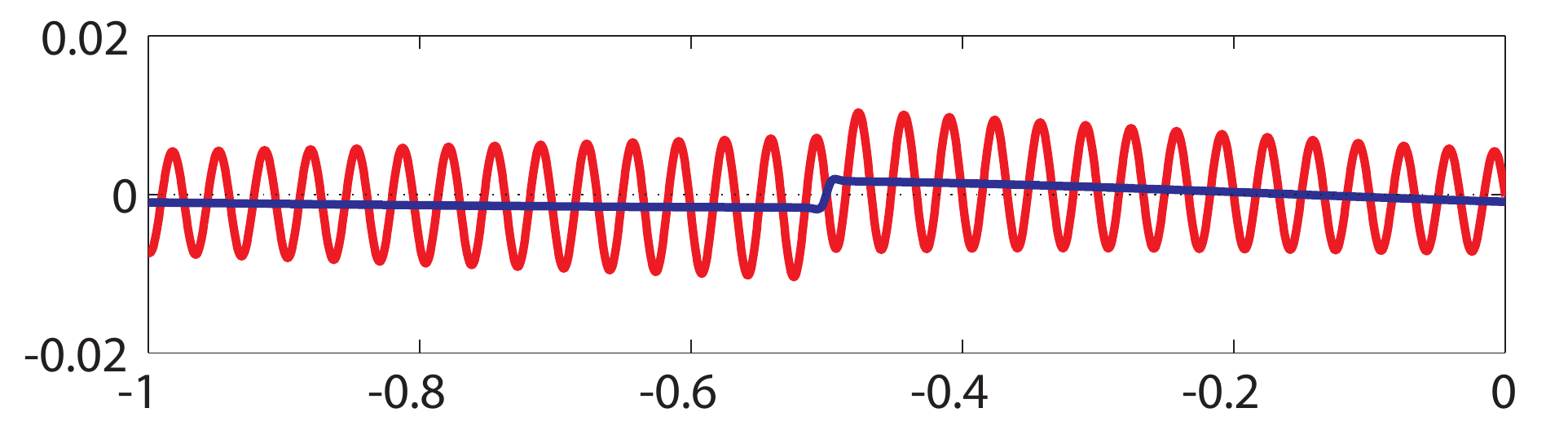} $x$\\
\vspace{-0.25cm}
(c) Two layer, $\alpha \approx 0.0018$, minus two layer control solution\\
\includegraphics[scale=0.65]{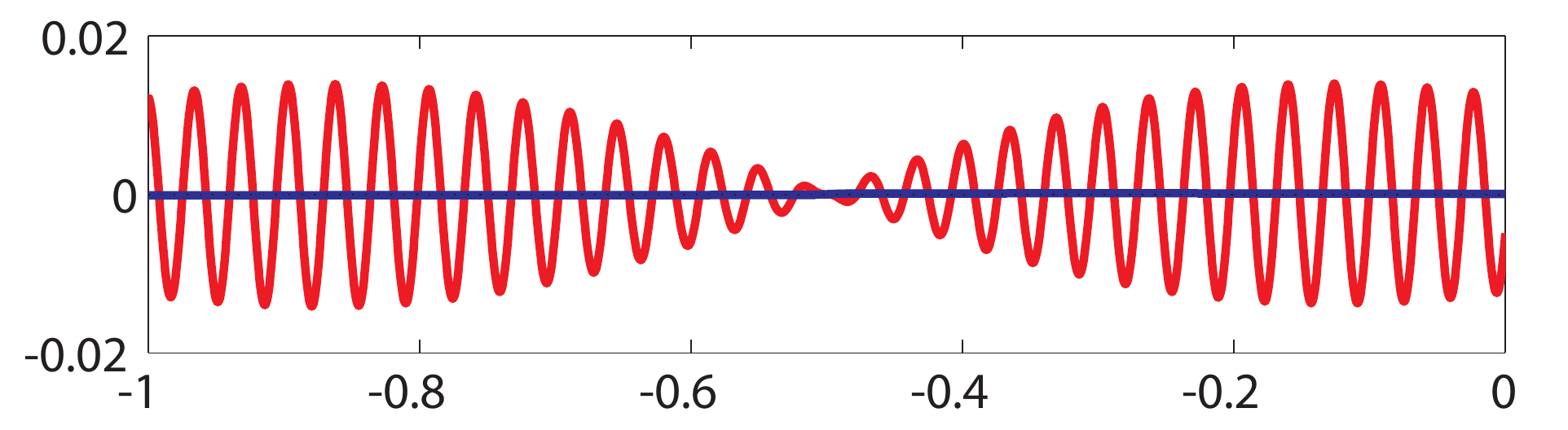} $x$\\
\vspace{-0.25cm}
(d) Two layer, $\gamma = 0.2$, minus two layer control solution\\
\includegraphics[scale=0.65]{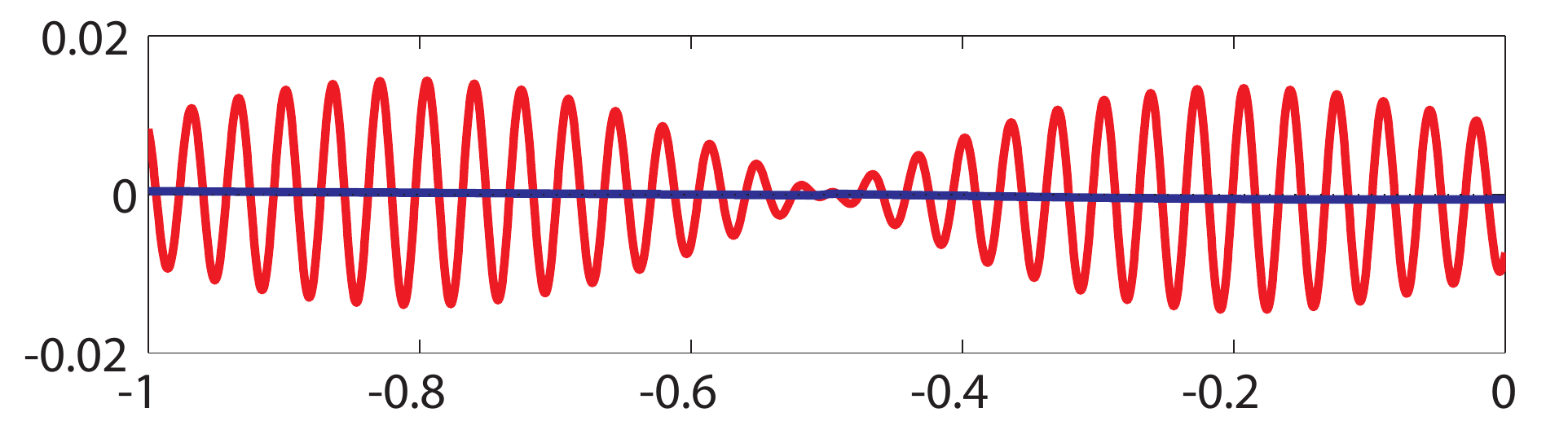}
$x$ \vspace{-0.5cm}
\caption{Surface elevation (red line) and barotropic surface elevation (blue line) amplitudes in an infinite basin. All plots were obtained using the Neumann series method. The $y$-axis in all plots is non-dimensional amplitude. The range in the vertical scales of (b), (c), and (d) is identical.} \label{fig:inf_magBT}
\end{center}
\end{figure}

\begin{figure}[h!]
\begin{center}
(a) One layer solution, $\alpha = 0$\\
\includegraphics[scale=0.65]{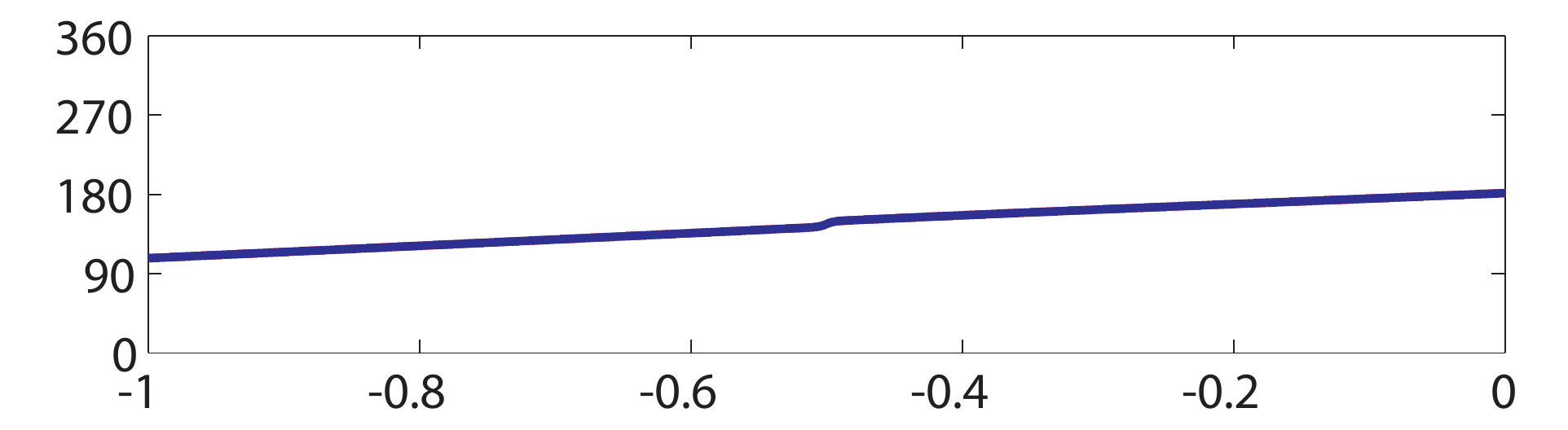} $x$\\
\vspace{-0.25cm}
(b) Two layer control, $\alpha \approx 0.0017$, minus one layer solution\\
\includegraphics[scale=0.65]{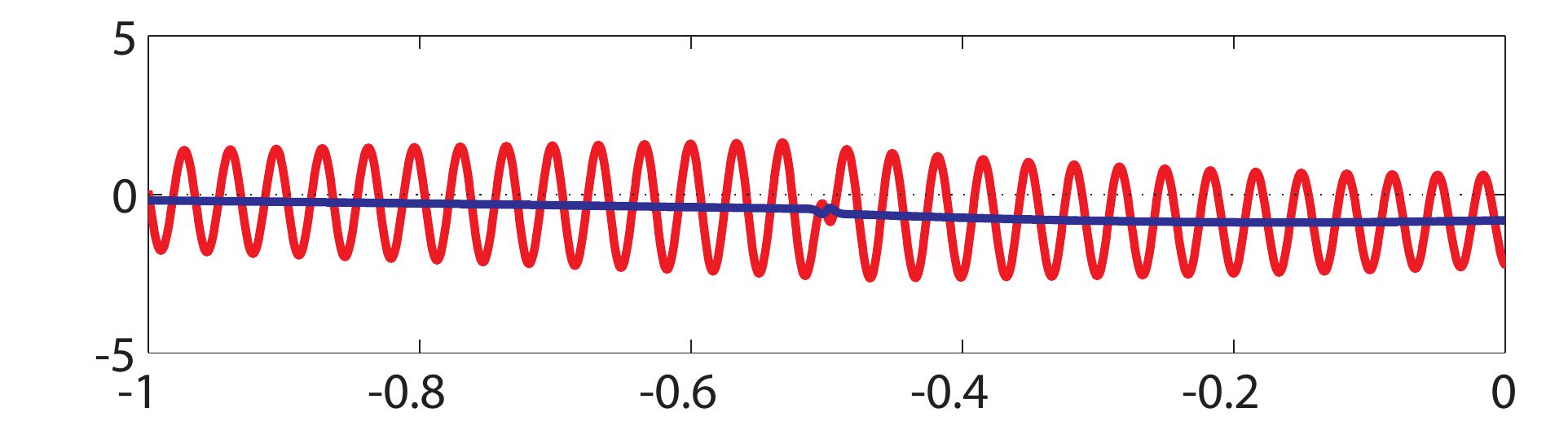} $x$\\
\vspace{-0.25cm}
(c) Two layer, $\alpha \approx 0.0018$, minus two layer control solution\\
\includegraphics[scale=0.65]{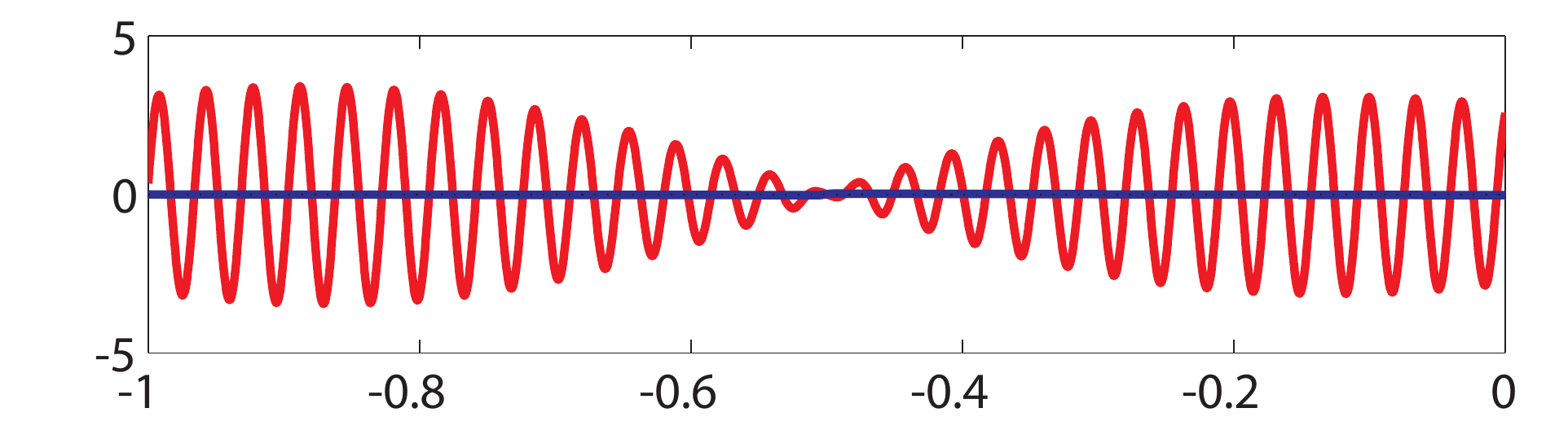} $x$\\
\vspace{-0.25cm}
(d) Two layer, $\gamma = 0.2$, minus two layer control solution\\
\includegraphics[scale=0.65]{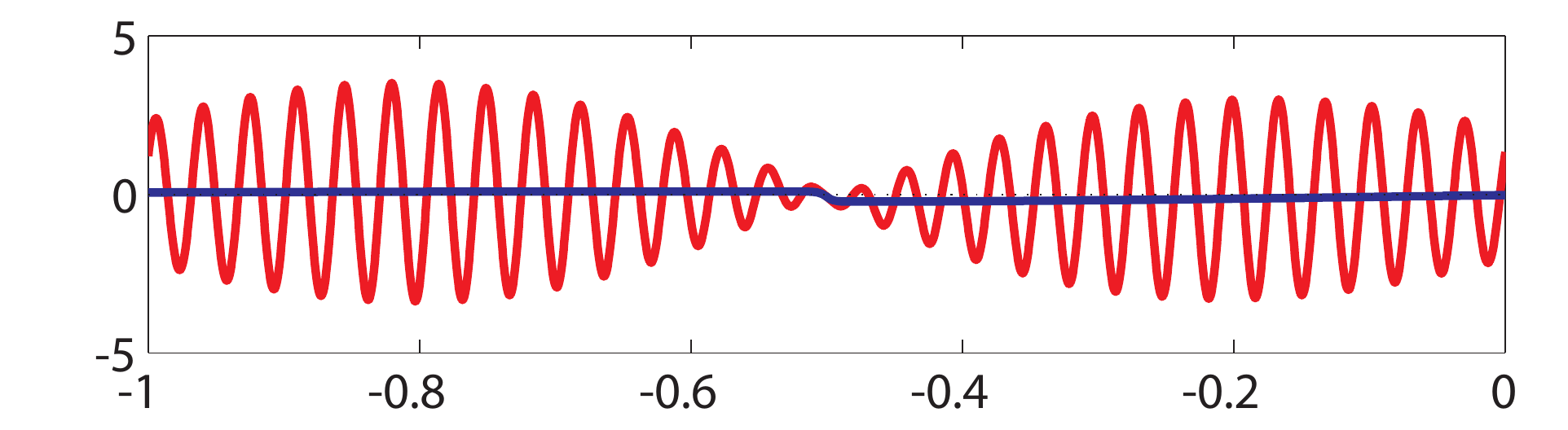}
$x$ \vspace{-0.5cm}
\caption{As in Figure \ref{fig:inf_magBT}, but for phase differences in degrees. The range in the vertical scales of (b), (c), and (d) is identical.} \label{fig:inf_phaBT}
\end{center}
\end{figure}

\begin{figure}[h!]
\begin{center}
(a) Two layer control, $\alpha \approx 0.0017$, minus one layer solution\\
\includegraphics[scale=0.65]{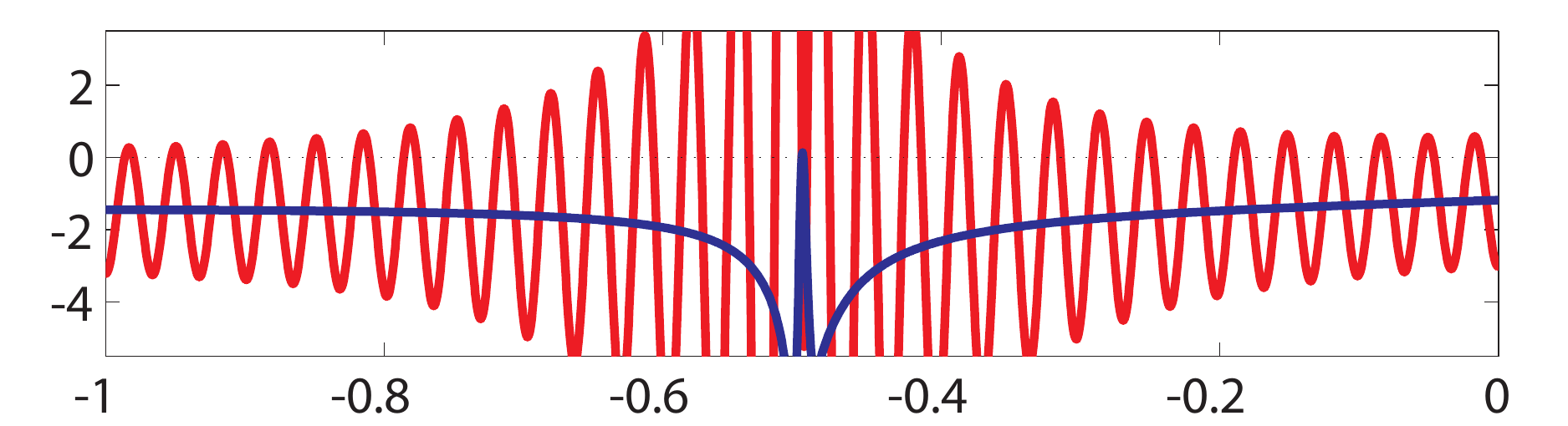} $x$\\
\vspace{-0.25cm}
(b) Two layer, $\alpha \approx 0.0018$, minus two layer control solution\\
\includegraphics[scale=0.65]{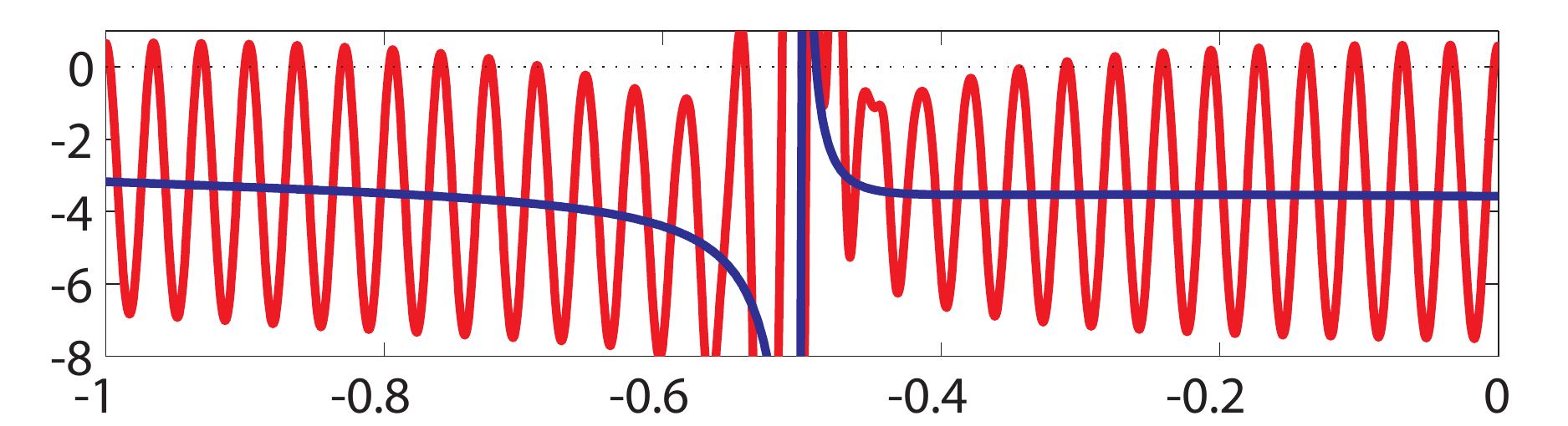} $x$\\
\vspace{-0.25cm}
(c) Two layer, $\gamma = 0.2$, minus two layer control solution\\
\includegraphics[scale=0.65]{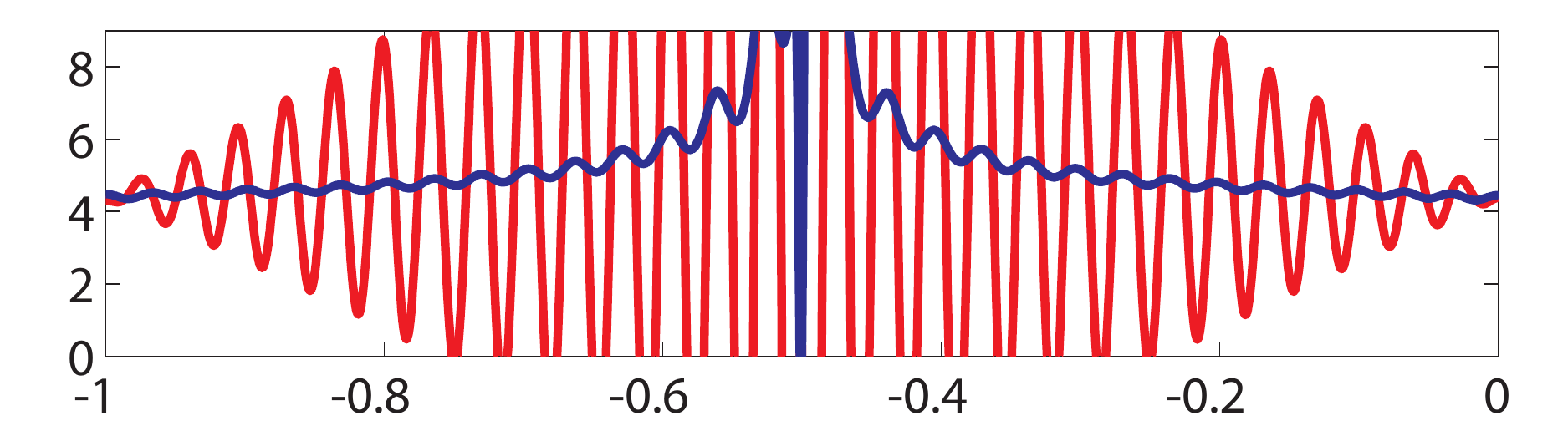}
$x$ \vspace{-0.5cm}
\caption{Surface elevation (red line) and barotropic surface elevation (blue line) amplitudes in a finite basin as a percentage of the control solution.  All plots were obtained using the Neumann series method. As in Figure \ref{fig:fin_phaBT} the large differences in the center of the basin are not as meaningful. The range in the vertical scales of all plots is identical.} \label{fig:fin_percent}
\end{center}
\end{figure}

\begin{figure}[h!]
\begin{center}
(a) $U^{BT}_{RMS}$\\
\includegraphics[scale=0.65]{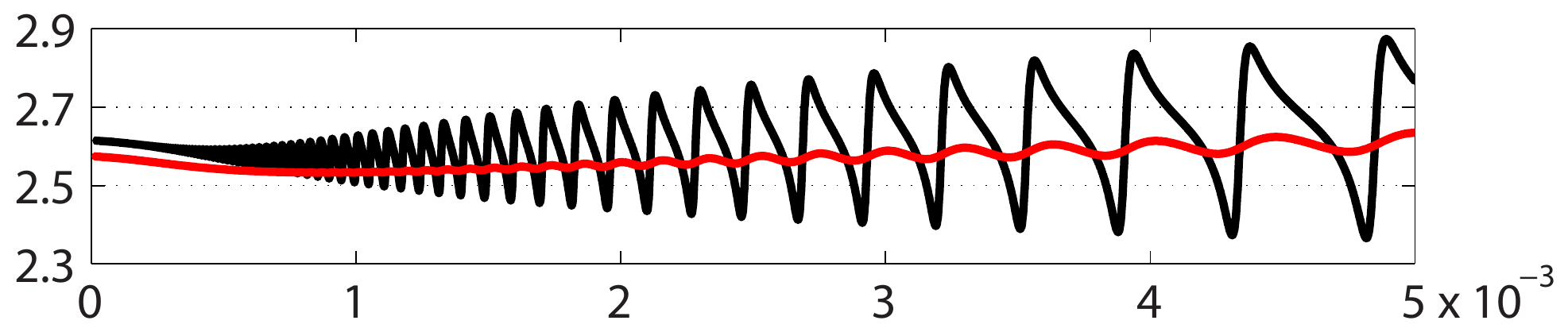} $\alpha$\\
\vspace{-0.25cm}
(b) $U^{BC}_{RMS}$\\
\includegraphics[scale=0.65]{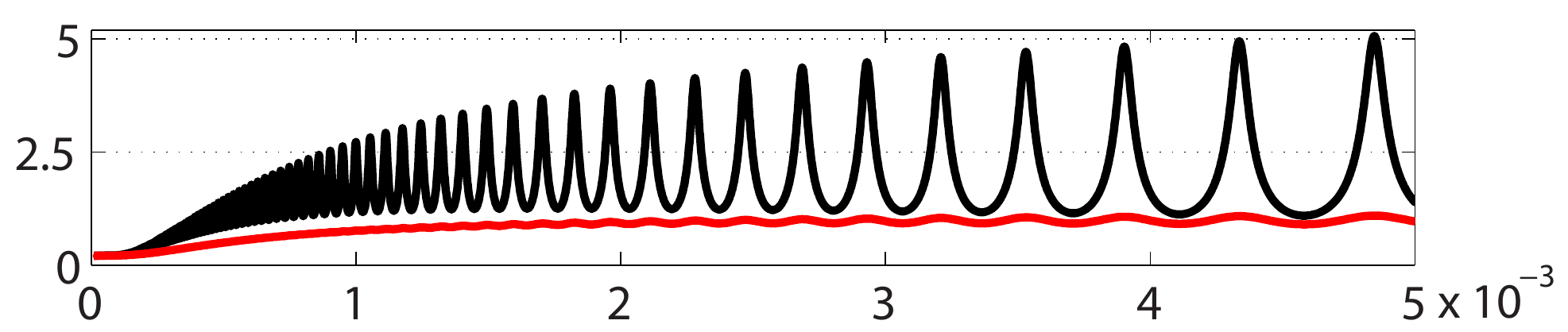} $\alpha$\\
\vspace{-0.25cm}
(c) $N_{1,RMS}^{BT}$\\
\includegraphics[scale=0.65]{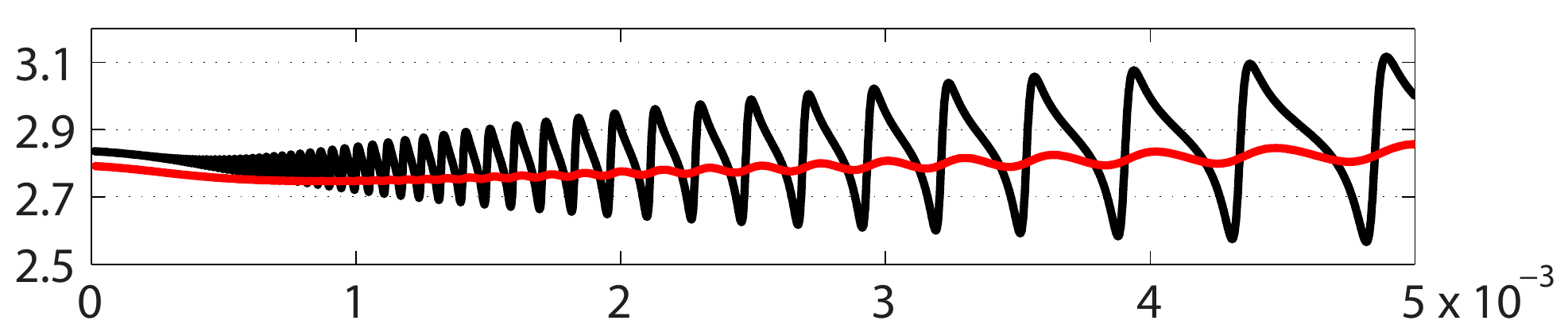} $\alpha$\\
\vspace{-0.25cm}
(d) $N_{1,RMS}^{BC}$\\
\includegraphics[scale=0.65]{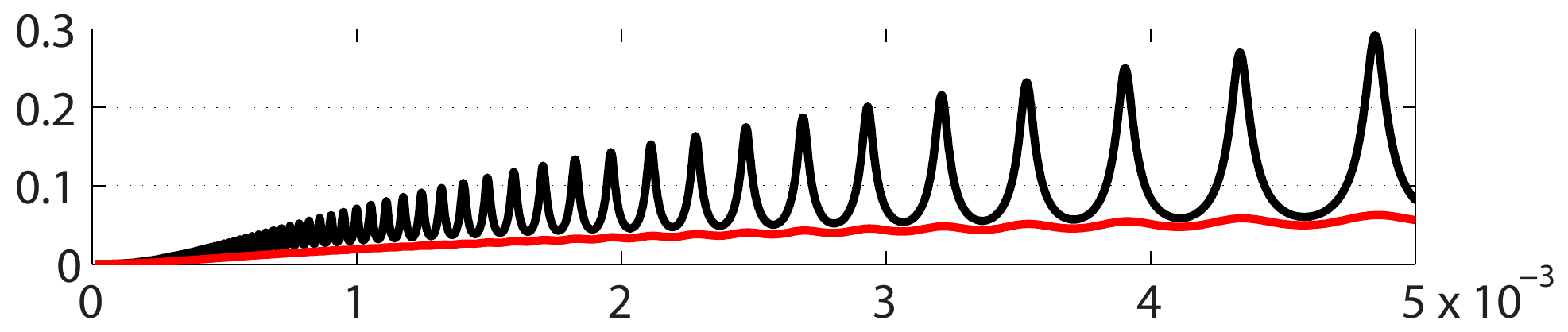}
$\alpha$ \vspace{-0.5cm}
\caption{RMS values versus stratification parameter $\alpha$ for two layers damped ($\delta_1 = \delta_2 \neq 0$), in red, and only the bottom layer damped ($\delta_1 =0, \delta_2 \neq 0$), in black.  All plots were obtained using the Fourier series method.} \label{fig:ene}
\end{center}
\end{figure}

\begin{figure}[h!]
\begin{center}
\includegraphics[scale=0.8]{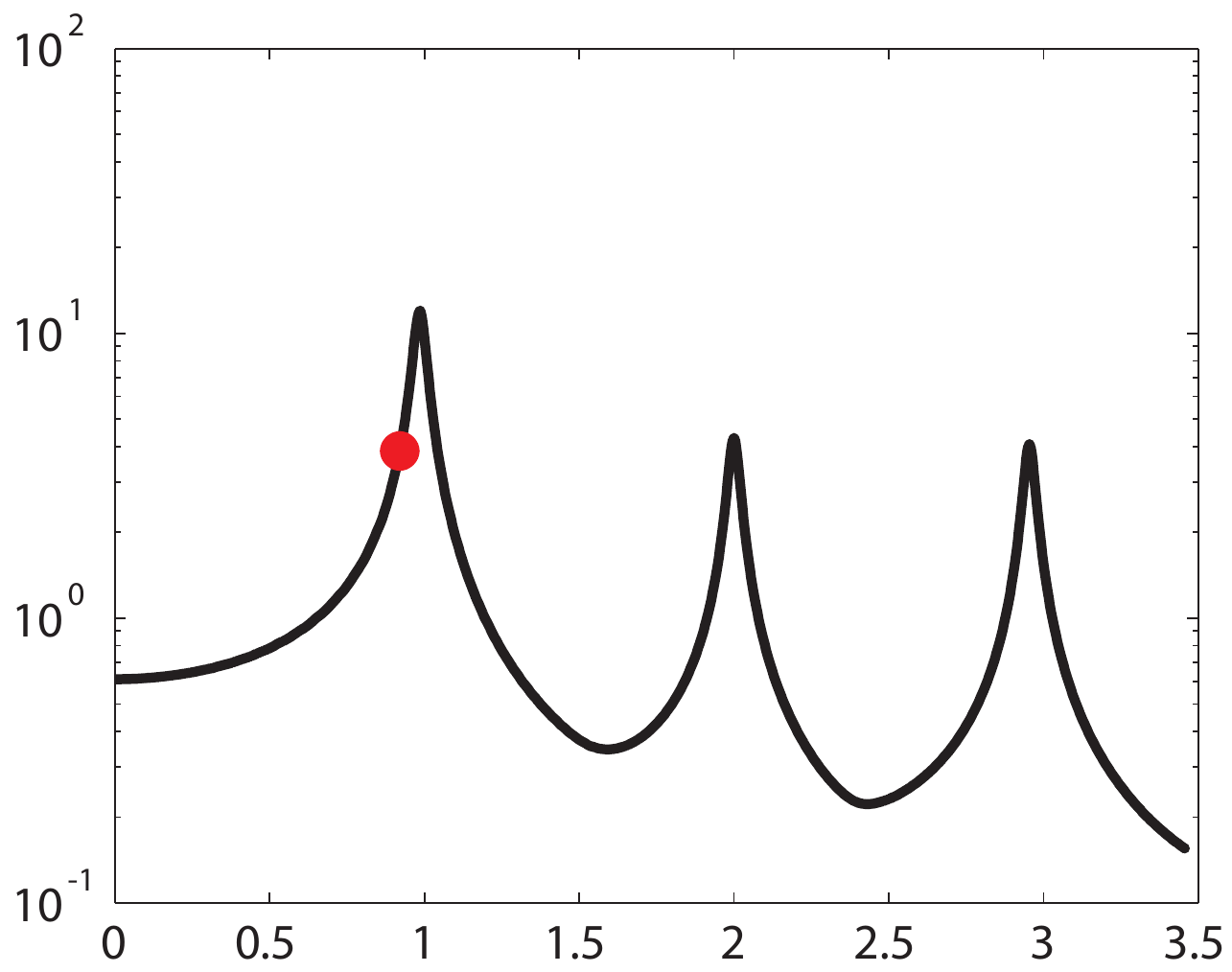}\\
\vspace{-8.5cm} \hspace{-12cm} $\max |N_1|$\\
\vspace{6.5cm} \hspace{11cm} $\displaystyle \frac{1}{\epsilon \pi}$
\caption{Maximum value of the surface elevation amplitude $N_1$ versus non-dimensional forcing frequency for the Gaussian topography described in section \ref{sec:num}. Red dot indicates position in parameter space for the control solution. Plot obtained using the Fourier series method.} \label{fig:resonan}
\end{center}
\end{figure}

For an infinite basin, changes to the stratification parameters may cause changes to the solution's amplitude as high as $10 \%$, see Figure \ref{fig:inf_magBT}, while changes in the phase of the solution are up to 4 degrees, see Figure \ref{fig:inf_phaBT}. As opposed to the finite basin, quite noticeable in all amplitude plots is the apparent lack of a measurable effect on the barotropic mode. Closer inspection shows that the barotropic mode indeed changes, but these changes are an order of magnitude smaller than those in the baroclinic mode's amplitude and hence are not easily observable; see Figure \ref{fig:inf_magBT} (b). Lastly, changes brought about by perturbing either of the stratification parameters, $\alpha$ or $\gamma$, may lead to qualitatively similar responses in the system's amplitude; compare Figures \ref{fig:inf_magBT} (c) and (d). 

A more global view of the sensitivity of the solution to the stratification parameter $\alpha$ can be obtained by computing the RMS value of relevant quantities as functions of $\alpha$. Here we define the RMS value as
\begin{equation}
\varphi_{RMS} := \sqrt{ \frac{1}{\ell_2-\ell_1} \int_{\ell_1}^{\ell_2} |\varphi(x)|^2 \; dx}.
\end{equation}
Figure \ref{fig:ene} shows the RMS values -- using the same parameter values as our two layer control -- for four variables: $U^{BT}$, $U^{BC}$, $N_1^{BT}$, and $N^{BC}_1$. The plots show strong sensitivity in the form of oscillations of significant amplitude reinforcing the fact that the stratification affects the surface barotropic mode. In particular, for $\alpha \approx 0.0017$ -- the $\alpha$ value of the two-layer control solution -- we see that the barotropic RMS quantities may be perturbed as much as $10\%$; see black curves in Figures \ref{fig:ene} (a) and (c). That is, the barotropic RMS values may change by as much as $10\%$ from peak to trough in the parameter range explored in this paper. It is worth noting that the observed oscillations are not about a constant mean; instead, the mean values in these plots vary with the stratification in an observable manner. This implies that these are not just oscillations about the limiting one-layer case. It is also observed that the sensitivity of the system to stratification is more significant when only the bottom layer is damped (black curves) than when both the top and bottom layers are identically damped (red curves).
The greater sensitivity in the black versus red curves implies that the tidal sensitivity to stratification is itself sensitive to assumptions about how tidal energy is dissipated.

Some understanding of the sensitivity of this system to stratification can be obtained from a resonance plot in the same vein as those in \cite{Arbic+al2009}. As in that paper, the quantity $\frac{1}{\epsilon \pi}$ acts as a non-dimensional measure of the forcing frequency of the system with higher values equivalent to faster forcing. Figure \ref{fig:resonan} shows that, for our chosen parameters, the control solution lies near a resonance peak of the system.
That is, the plot shows that effects of stratification on the barotropic surface tidal elevation might be accentuated due to our chosen location in parameter space.

%%%%%%%%%%%%%%%%%%

\section{Summary}

We have presented an analytical tide model which demonstrates the impacts of stratification on surface tidal elevations in a qualitatively similar manner as in the global numerical simulations shown in \cite{Arbic+al2004}, \cite{Shriver+al2014}, and section \ref{sec:numerical}. 
Our analytical and numerical results demonstrate the potential for the presence of stratification to impact the barotropic tide, and for climatic perturbations of oceanic stratifications to contribute to the secular changes in tides observed in tide gauge records. 
Our derived analytical formulas for the barotropic and baroclinic modal contributions to surface tidal elevations contain explicit dependence on the stratification parameters $\alpha$ and $\gamma$, implying that the barotropic as well as the baroclinic mode is affected by stratification.
This dependence can be understood, at a basic level, by the fact that changes in stratification affect the speed of barotropic gravity wave propagation in the basin which in turn change the tidal amplitude. We quantify these effects with surface elevation plots in both finite and infinite basins over a Gaussian bump and plots of the root mean square values of model velocities and elevations using representative oceanic parameters. We find that, for the given parameters, our finite basin model's barotropic mode is much more affected by stratification than that of our infinite domain model. A significant contribution to this effect arises from the size of the topography and particularly the nearness of the basin boundaries.
For the finite basin, changes in stratification change the barotropic tide by as much as $10\%$, making these perturbations of a size comparable to the baroclinic signal of the tide.

%%%%%%%%%%%%%%%%%%%%%%%%%%%%%%%%%%%%%%%%%%%%%%%%%%%%%%%%%%%%%%%%%%%%%
% ACKNOWLEDGMENTS
%%%%%%%%%%%%%%%%%%%%%%%%%%%%%%%%%%%%%%%%%%%%%%%%%%%%%%%%%%%%%%%%%%%%%
%
\acknowledgments

BKA and ANW gratefully acknowledge support from National Science Foundation (NSF) grants OCE-0924481 and OCE-0960820.
IC and MCH gratefully acknowledge support from the Department of Boating and Waterways of the State of California, grant 08-106-105. PDM and ANW were partially supported by the NSF under grant DMS-1206131.
JFM was supported by a Research Experience for Undergraduates (REU) supplement to OCE-0924481.
We acknowledge Detlef Stammer for organizing a workshop on secular changes in tides, which provided additional inspiration for this study.
Useful discussions with Peter Brominski, Bill Dewar, Gary Egbert, Ron Flick, Chris Garrett, Malte M\"uller, and  Richard Ray are gratefully acknowledged.
Numerical computations were performed on the High-Performance Computing (HPC) centers at Florida State University.
BKA thanks Jordan Yao, Dan Voss, and Paul van der Mark for technical support on the HPC. \nocite{Arbic+al2004} \cite{Arbic+al2009} \cite{Arbic+Owens2001}

%%%%%%%%%%%%%%%%%%%%%%%%%%%%%%%%%%%%%%%%%%%%%%%%%%%%%%%%%%%%%%%%%%%%%
% APPENDIXES
%%%%%%%%%%%%%%%%%%%%%%%%%%%%%%%%%%%%%%%%%%%%%%%%%%%%%%%%%%%%%%%%%%%%%
%
% Use \appendix if there is only one appendix.
%\appendix

% Use \appendix[A], \appendix}[B], if you have multiple appendixes.
%\appendix[A]

%% Appendix title is necessary! For appendix title:
%\appendixtitle{}

%%% Appendix section numbering (note, skip \section and begin with \subsection)
% \subsection{First primary heading}

% \subsubsection{First secondary heading}

% \paragraph{First tertiary heading}

%% Important!
%\appendcaption{<appendix letter and number>}{<caption>} 
%must be used for figures and tables in appendixes, e.g.,
%
%\begin{figure}
%\noindent\includegraphics[width=19pc,angle=0]{figure01.pdf}\\
%\appendcaption{A1}{Caption here.}
%\end{figure}

%%%%%%%%%%%%%%%%%%%%%%%%%%%%%%%%%%%%%%%%%%%%%%%%%%%%%%%%%%%%%%%%%%%%%
% REFERENCES
%%%%%%%%%%%%%%%%%%%%%%%%%%%%%%%%%%%%%%%%%%%%%%%%%%%%%%%%%%%%%%%%%%%%%
% Make your BibTeX bibliography by using these commands:
\bibliographystyle{ametsoc2014}
\bibliography{references}

\begin{thebibliography}{32}
\providecommand{\natexlab}[1]{#1}
\providecommand{\url}[1]{\texttt{#1}}
\renewcommand{\UrlFont}{\rmfamily}
\providecommand{\urlprefix}{URL }
\expandafter\ifx\csname urlstyle\endcsname\relax
  \providecommand{\doi}[1]{doi:\discretionary{}{}{}#1}\else
  \providecommand{\doi}{doi:\discretionary{}{}{}\begingroup
  \urlstyle{rm}\Url}\fi
\providecommand{\eprint}[2][]{\url{#2}}

\bibitem[{Arbic et~al.(2004)Arbic, Garner, Hallberg,, and
  Simmons}]{Arbic+al2004}
Arbic, B.~K., S.~T. Garner, R.~W. Hallberg, and H.~L. Simmons, 2004: The
  accuracy of surface elevations in forward global barotropic and baroclinic
  tide models. \textit{Deep-Sea Res. II}, \textbf{51}, 3069--3101,
  \doi{10.1016/j.dsr2.2004.09.014}.

\bibitem[{Arbic et~al.(2009)Arbic, Karsten,, and Garrett}]{Arbic+al2009}
Arbic, B.~K., R.~H. Karsten, and C.~Garrett, 2009: On tidal resonance in the
  global ocean and the back-effect of coastal tides upon open-ocean tides.
  \textit{Atmosphere-Ocean}, \textbf{47}, 239--266, \doi{10.3137/OC311.2009}.

\bibitem[{Arbic and Owens(2001)Arbic, and Owens}]{Arbic+Owens2001}
Arbic, B.~K., and W.~B. Owens, 2001: Climatic warming of {A}tlantic
  intermediate waters. \textit{J. Clim.}, \textbf{14}, 4091--4108,
  \doi{10.1175/1520-0442(2001)014<4091:CWOAIW>2.0.CO;2}.

\bibitem[{Arbic et~al.(2010)Arbic, Wallcraft,, and Metzger}]{Arbic+al2010}
Arbic, B.~K., A.~J. Wallcraft, and E.~J. Metzger, 2010: Concurrent simulation
  of the eddying general circulation and tides in a global ocean model.
  \textit{Ocean Modelling}, \textbf{32}, 175--187,
  \doi{10.1016/j.ocemod.2010.01.007}.

\bibitem[{Cartwright(1971)}]{Cartwright1971}
Cartwright, D.~E., 1971: Tides and waves in the vicinity of {S}aint {H}elena.
  \textit{Phil. Trans. R. Soc. Lond. A}, \textbf{270}, 603--646,
  \doi{10.1098/rsta.1971.0091}.

\bibitem[{Colosi and Munk(2006)Colosi, and Munk}]{Colosi+Munk2006}
Colosi, J.~A., and W.~Munk, 2006: Tales of the venerable {H}onolulu tide gauge.
  \textit{J. Phys. Oceanogr.}, \textbf{36}, 967--996, \doi{10.1175/JPO2876.1}.

\bibitem[{Egbert et~al.(2004)Egbert, Ray,, and Bills}]{Egbert+al2004}
Egbert, G.~D., R.~D. Ray, and B.~G. Bills, 2004: Numerical modeling of the
  global semidiurnal tide in the present day and in the last glacial maximum.
  \textit{J. Geophys. Res.}, \textbf{109~(C3)}, \doi{10.1029/2003JC001973}.

\bibitem[{Flick et~al.(2003)Flick, Murray,, and Ewing}]{Flick+al2003}
Flick, R.~E., J.~F. Murray, and L.~C. Ewing, 2003: Trends in united states
  datum statistics and tide range. \textit{J. Waterway, Port, Coastal and Ocean
  Engin.}, \textbf{129}, 155--164,
  \doi{10.1061/(ASCE)0733-950X(2003)129:4(155)}.

\bibitem[{Flierl(1978)}]{Flierl1978}
Flierl, G.~R., 1978: Models of vertical structure and the calibration of
  two-layer models. \textit{Dyn. Atmos. Oceans}, \textbf{2}, 341--381,
  \doi{10.1016/0377-0265(78)90002-7}.

\bibitem[{Gill(1982)}]{Gill1982}
Gill, A.~E., 1982: \textit{Atmosphere-Ocean Dynamics}. Academic Press, 662 pp.

\bibitem[{Greenberg(1971)}]{Greenberg1971}
Greenberg, M.~D., 1971: \textit{Application of Green's Functions in Science and
  Engineering}. Prentice-Hall Inc., 141 pp.

\bibitem[{Hallberg and Rhines(1996)Hallberg, and Rhines}]{Hallberg+Rhines1996}
Hallberg, R., and P.~Rhines, 1996: Boundary-driven circulation in an ocean
  basin with isopycnals intersecting the sloping boundary. \textit{J. Phys.
  Oceanogr.}, \textbf{26}, 914--940,
  \doi{10.1175/1520-0485(1996)026<0913:BDCIAO>2.0.CO;2}.

\bibitem[{Hendershott(1972)}]{Hendershott1972}
Hendershott, M.~C., 1972: The effects of solid earth deformation on global
  ocean tides. \textit{Geophys. J. Royal Astr. Soc.}, \textbf{29}, 389--402,
  \doi{10.1111/j.1365-246X.1972.tb06167.x}.

\bibitem[{Jay(2009)}]{Jay2009}
Jay, D.~A., 2009: Evolution of tidal amplitudes in the eastern pacific ocean.
  \textit{Geophys. Res. Lett.}, \textbf{36}, L04\,603,
  \doi{10.1029/2008GL036185}.

\bibitem[{Jayne and {St. Laurent}(2001)Jayne, and {St.
  Laurent}}]{Jayne+StLaurent2001}
Jayne, S.~R., and L.~C. {St. Laurent}, 2001: Parametrizing tidal dissipation
  over rough topography. \textit{Geophys. Res. Lett.}, \textbf{28}, 811--814,
  \doi{10.1029/2000GL012044}.

\bibitem[{Kang et~al.(2002)Kang, Foreman, Lie, Lee, Cherniawsky,, and
  Yum}]{Kang+al2002}
Kang, S.~K., M.~G.~G. Foreman, H.-J. Lie, J.-H. Lee, J.~Cherniawsky, and K.-D.
  Yum, 2002: Two-layer tide modeling of the {Y}ellow and {E}ast {C}hina seas
  with application to seasonal variability of the {$M_{2}$} tide. \textit{J.
  Geophys. Res.}, \textbf{107}, 6--1--6--18, \doi{10.1029/2001JC000838}.

\bibitem[{Khatiwala(2003)}]{Khatiwala2003}
Khatiwala, S., 2003: Generation of internal tides in an ocean of finite depth:
  analytical and numerical calculations. \textit{Dee-Sea Res. I}, \textbf{51},
  3--21, \doi{10.1016/S0967-0637(02)00132-2}.

\bibitem[{{Llewellyn Smith} and Young(2001){Llewellyn Smith}, and
  Young}]{Llewellyn+al2001}
{Llewellyn Smith}, S.~G., and W.~R. Young, 2001: Conversion of the barotropic
  tide. \textit{J. Phys. Oceanogr.}, \textbf{32}, 1554--1566,
  \doi{10.1175/1520-0485(2002)032<1554:COTBT>2.0.CO;2}.

\bibitem[{M\"uller(2012)}]{Muller2012}
M\"uller, M., 2012: The influence of changing stratification conditions on
  barotropic tidal transport and its implications for seasonal and secular
  changes of tides. \textit{Cont. Shelf Res.}, \textbf{47}, 107--118,
  \doi{10.1016/j.csr.2012.07.003}.

\bibitem[{M\"uller et~al.(2011)M\"uller, Arbic,, and Mitrovica}]{Muller+al2011}
M\"uller, M., B.~K. Arbic, and J.~X. Mitrovica, 2011: Secular changes in ocean
  tides: Observations and model results. \textit{J. Geophys. Res.},
  \textbf{116}, C05\,013, \doi{10.1029/2010JC006387}.

\bibitem[{M\"uller(2006)}]{Muller2006}
M\"uller, P., 2006: \textit{The Equations of Oceanic Motions}. Cambridge
  University Press, 291 pp.

\bibitem[{Munk(1997)}]{Munk1997}
Munk, W., 1997: Once again: Once again - tidal friction. \textit{Progress in
  Oceanography}, \textbf{40}, 7--35, \doi{10.1016/S0079-6611(97)00021-9}.

\bibitem[{Ray(1999)}]{Ray1999}
Ray, R.~D., 1999: A global ocean tide model from {TOPEX/POSEIDON} altimetry:
  {GOT}99.2, {N}ational {A}eronautics and {S}pace {A}dministration technical
  memorandum. Tech. rep. NASA/TM-1999-209478.

\bibitem[{Ray(2006)}]{Ray2006}
Ray, R.~D., 2006: Secular changes of the {$M_2$} tide in the gulf of maine.
  \textit{Continental Shelf Res.}, \textbf{26}, 422--427,
  \doi{10.1016/j.csr.2005.12.005}.

\bibitem[{Ray(2009)}]{Ray2009}
Ray, R.~D., 2009: Secular changes in the solar semidiurnal tide of the western
  {N}orth {A}tlantic {O}cean. \textit{Geophys. Res. Lett.}, \textbf{36},
  L19\,601, \doi{10.1029/2009GL040217}.

\bibitem[{Ray and Mitchum(1996)Ray, and Mitchum}]{Ray+Mitchum1996}
Ray, R.~D., and G.~T. Mitchum, 1996: Surface manifestation of internal tides
  generated near {H}awaii. \textit{Geophys. Res. Lett.}, \textbf{23},
  2101--2104, \doi{10.1029/96GL02050}.

\bibitem[{Ray and Mitchum(1997)Ray, and Mitchum}]{Ray+Mitchum1997}
Ray, R.~D., and G.~T. Mitchum, 1997: Surface manifestation of internal tides in
  the deep ocean: observations from altimetry and island gauges.
  \textit{Progress in Oceanography}, \textbf{40}, 135--162,
  \doi{10.1016/S0079-6611(97)00025-6}.

\bibitem[{Shriver et~al.(2014)Shriver, Richman,, and Arbic}]{Shriver+al2014}
Shriver, J.~F., J.~G. Richman, and B.~K. Arbic, 2014: How stationary are the
  internal tides in a high-resolution global ocean circulation model?
  \textit{J. Geophys. Res.}, \textbf{119}, 2769--2787,
  \doi{10.1002/2013JC009423}.

\bibitem[{{St. Laurent} et~al.(2003){St. Laurent}, Stringer, Garrett,, and
  Perrault-Joncas}]{StLaurent+al2003}
{St. Laurent}, L.~C., S.~Stringer, C.~Garrett, and D.~Perrault-Joncas, 2003:
  The generation of internal tides at abrupt topography. \textit{Deep-Sea Res.
  I}, \textbf{50}, 987--1003, \doi{10.1016/S0967-0637(03)00096-7}.

\bibitem[{Woodworth(2010)}]{Woodworth2010}
Woodworth, P.~L., 2010: A survey of recent changes in the main components of
  the ocean tide. \textit{Continental Shelf Res.}, \textbf{30}, 1680--1691,
  \doi{10.1016/j.csr.2010.07.002}.

\bibitem[{Woodworth et~al.(1991)Woodworth, Shaw,, and
  Blackman}]{Woodworth+al1991}
Woodworth, P.~L., S.~M. Shaw, and D.~L. Blackman, 1991: Secular trends in mean
  tidal range around the {B}ritish {I}sles and along the adjacent {E}uropean
  coastline. \textit{Geophys. J. Int.}, \textbf{104}, 593--609,
  \doi{10.1111/j.1365-246X.1991.tb05704.x}.

\bibitem[{Zaron and Jay(2014)Zaron, and Jay}]{Zaron+Jay2014}
Zaron, E.~D., and D.~A. Jay, 2014: An analysis of secular change in tides at
  open-ocean sites in the pacific. \textit{J. Phys. Oceanogr.},
  \doi{10.1175/JPO-D-13-0266.1}, in press.

\end{thebibliography}

%%%%%%%%%%%%%%%%%%%%%%%%%%%%%%%%%%%%%%%%%%%%%%%%%%%%%%%%%%%%%%%%%%%%%
% TABLES
%%%%%%%%%%%%%%%%%%%%%%%%%%%%%%%%%%%%%%%%%%%%%%%%%%%%%%%%%%%%%%%%%%%%%
%% Enter tables at the end of the document, before figures.
%%
%
%\begin{table}[t]
%\caption{This is a sample table caption and table layout.  Enter as many tables as
%  necessary at the end of your manuscript. Table from Lorenz (1963).}\label{t1}
%\begin{center}
%\begin{tabular}{ccccrrcrc}
%\hline\hline
%$N$ & $X$ & $Y$ & $Z$\\
%\hline
% 0000 & 0000 & 0010 & 0000 \\
% 0005 & 0004 & 0012 & 0000 \\
% 0010 & 0009 & 0020 & 0000 \\
% 0015 & 0016 & 0036 & 0002 \\
% 0020 & 0030 & 0066 & 0007 \\
% 0025 & 0054 & 0115 & 0024 \\
%\hline
%\end{tabular}
%\end{center}
%\end{table}

%%%%%%%%%%%%%%%%%%%%%%%%%%%%%%%%%%%%%%%%%%%%%%%%%%%%%%%%%%%%%%%%%%%%%
% FIGURES
%%%%%%%%%%%%%%%%%%%%%%%%%%%%%%%%%%%%%%%%%%%%%%%%%%%%%%%%%%%%%%%%%%%%%
%% Enter figures at the end of the document, after tables.
%%
%
%\listoffigures

\newpage

\newpage

\newpage

\newpage

\newpage

\newpage

\newpage

\newpage

\newpage

\end{document}